\documentclass[11pt]{article}
\usepackage[pdftex]{graphicx}
\usepackage[lmargin=1.5in, rmargin=1.5in, tmargin=1.25in, bmargin=1.25in]{geometry}

\usepackage{amsmath}
\usepackage{amssymb}
\usepackage{amsthm}
\usepackage{mathtools}
\usepackage{mathdots}
\usepackage{fancyhdr}
\usepackage{multicol}
\usepackage{bm}
\usepackage{listings}
\PassOptionsToPackage{usenames,dvipsnames}{color}  
\usepackage{pdfpages}
\usepackage{algpseudocode}
\usepackage{tikz}
\usepackage[T1]{fontenc}
\usepackage{inconsolata}
\usepackage{framed}
\usepackage{wasysym}
\usepackage[thinlines]{easytable}
\usepackage{hyperref}
\hypersetup{colorlinks,citecolor=blue,urlcolor=blue,linkcolor=blue}
\usepackage[ruled]{algorithm2e}
\usepackage[round, comma, authoryear, longnamesfirst]{natbib}
\usepackage[nodisplayskipstretch]{setspace}
\usepackage{float}
\allowdisplaybreaks
\usepackage{etoolbox}
\makeatletter
\patchcmd{\Ginclude@eps}{"#1"}{#1}{}{}
\makeatother

\newcommand{\abs}[1]{\lvert #1 \rvert}
\newcommand{\absfit}[1]{\left\lvert #1 \right\rvert}
\newcommand{\normnofit}[1]{\lVert #1 \rVert}
\newcommand{\norm}[1]{\left\lVert #1 \right\rVert}

\newcommand{\ceil}[1]{\left\lceil#1\right\rceil}

\newcommand{\ind}[1]{\mathbf{1}\left\{#1\right\}}

\newcommand{\bmat}[1]{\begin{bmatrix}#1\end{bmatrix}}

\DeclareMathOperator{\R}{\mathbb{R}}

\DeclareMathOperator*{\argmin}{arg\,min}

\DeclareMathOperator{\N}{\mathbb{N}}

\DeclareMathOperator{\zeros}{\mathbf{0}}
\DeclareMathOperator{\ones}{\mathbf{1}}
\let\epsilon\relax
\DeclareMathOperator{\epsilon}{\varepsilon}
\DeclareMathOperator{\setst}{:}
\DeclareMathOperator{\suchthat}{s.t.}

\definecolor{shadecolor}{gray}{0.95}

\usepackage{enumitem}
\usepackage[font=footnotesize, skip=8pt]{caption}
\usepackage{subcaption}
\usepackage[bottom]{footmisc}
\usepackage{array}
\usepackage{multirow}

\theoremstyle{definition}

\newfloat{algbox}{tbp}{lop}
\newtheorem{alg}{Procedure}
\newcommand{\myalg}[3]{
\begin{center}
\fbox{
\parbox{0.9\textwidth}{
\begin{alg}\label{#1}{\textsc{#2}}
#3
\end{alg}
}}
\end{center}
}

\title{Assessing the Sensitivity of Synthetic Control Treatment Effect Estimates to Misspecification Error\footnote{
	We are grateful for invaluable comments from and stimulating conversations with Bharat Chandar, Jiafeng Chen, Benny Goldman, Guido Imbens, Advik Shreekumar, Charlie Walker, and the participants in the Stanford Econometrics and Applied Lunches.
		}}
\author{Billy Ferguson \\ Graduate School of Business \\ Stanford University \\ \texttt{billyf@stanford.edu}
		\and Brad Ross \\ Graduate School of Business \\ Stanford University \\ \texttt{bradross@stanford.edu}}
\date{Last Updated \today}

\begin{document}

\maketitle

\begin{abstract}
We propose a sensitivity analysis for Synthetic Control (SC) treatment effect estimates to interrogate the assumption that the SC method is well-specified, namely that choosing weights to minimize pre-treatment prediction error yields accurate predictions of counterfactual post-treatment outcomes. Our data-driven procedure recovers the set of treatment effects consistent with the assumption that the misspecification error incurred by the SC method is at most the observable misspecification error incurred when using the SC estimator to predict the outcomes of some control unit. We show that under one definition of misspecification error, our procedure provides a simple, geometric motivation for comparing the estimated treatment effect to the distribution of placebo residuals to assess estimate credibility. When we apply our procedure to several canonical studies that report SC estimates, we broadly confirm the conclusions drawn by the source papers.
\end{abstract}

\section{Introduction}\label{sec:intro}

The Synthetic Control (SC) method was originally developed in \cite{abadie2003basque}, \cite{abadie2010synthetic}, and \cite{abadie2015comparative} to estimate treatment effects in comparative case study settings, in which a researcher observes panel data on aggregate outcomes for a small number of large, heterogeneous units, only one of which receives some intervention of interest at some point in time. For the SC method to yield credible treatment effect estimates for the treated unit, researchers must assume that the SC method is well-specified: if there is a convex combination of control units' pre-treatment outcomes that closely approximates the pre-treatment outcomes of the treated unit, then that same convex combination of control units' post-treatment outcomes will yield good estimates of the treated unit's post-treatment control outcomes.\footnote{In addition, the researcher must assume that there are no idiosyncratic factors that affect the treated unit's counterfactual control outcomes post-treatment but not the control units' post-treatment outcomes besides their differing treatment statuses; one way to operationalize this idea is the linear factor model presented in \cite{abadie2010synthetic} and studied in depth in \cite{ferman2019synthetic}, in which units' factor loadings do not vary before and after treatment. It is also important to assume that the treatment does not affect units in the donor pool, although researchers can simply exclude ``control'' units for which spillover effects are a concern \citep{abadie2020using}. Typically, such assumptions must be justified using domain knowledge about the setting of interest, so we do not concern ourselves with assessing their validity in this paper.} While this assumption is necessary for tractable treatment effect estimation, it is unlikely to hold exactly in practice. In this paper, we develop a sensitivity analysis of SC treatment effect estimates that bounds the true treatment effect under the assumption that any deviations from this well-specified method assumption are at most as severe as the deviations observed in placebo analyses of the control units. 

To build intuition for where this assumption might lead researchers astray, we present two placebo analyses in which we apply the SC method to panel data from the evaluation of a 1989 tobacco control program implemented in California, as in \citet{abadie2010synthetic}. In particular, we use the SC method to predict per-capita tobacco sales in Virginia and Delaware in the year 2000 using the other members of the donor pool as control units. Since neither Virginia nor Delaware received the treatment and we observe their true control outcomes post-treatment, we can see whether the SC method correctly predicts no effects in either state.

\begin{figure}[t]
	\captionsetup{width=0.93\textwidth}
	\begin{subfigure}{0.5\textwidth}
	\centering
	\includegraphics[width=\textwidth]{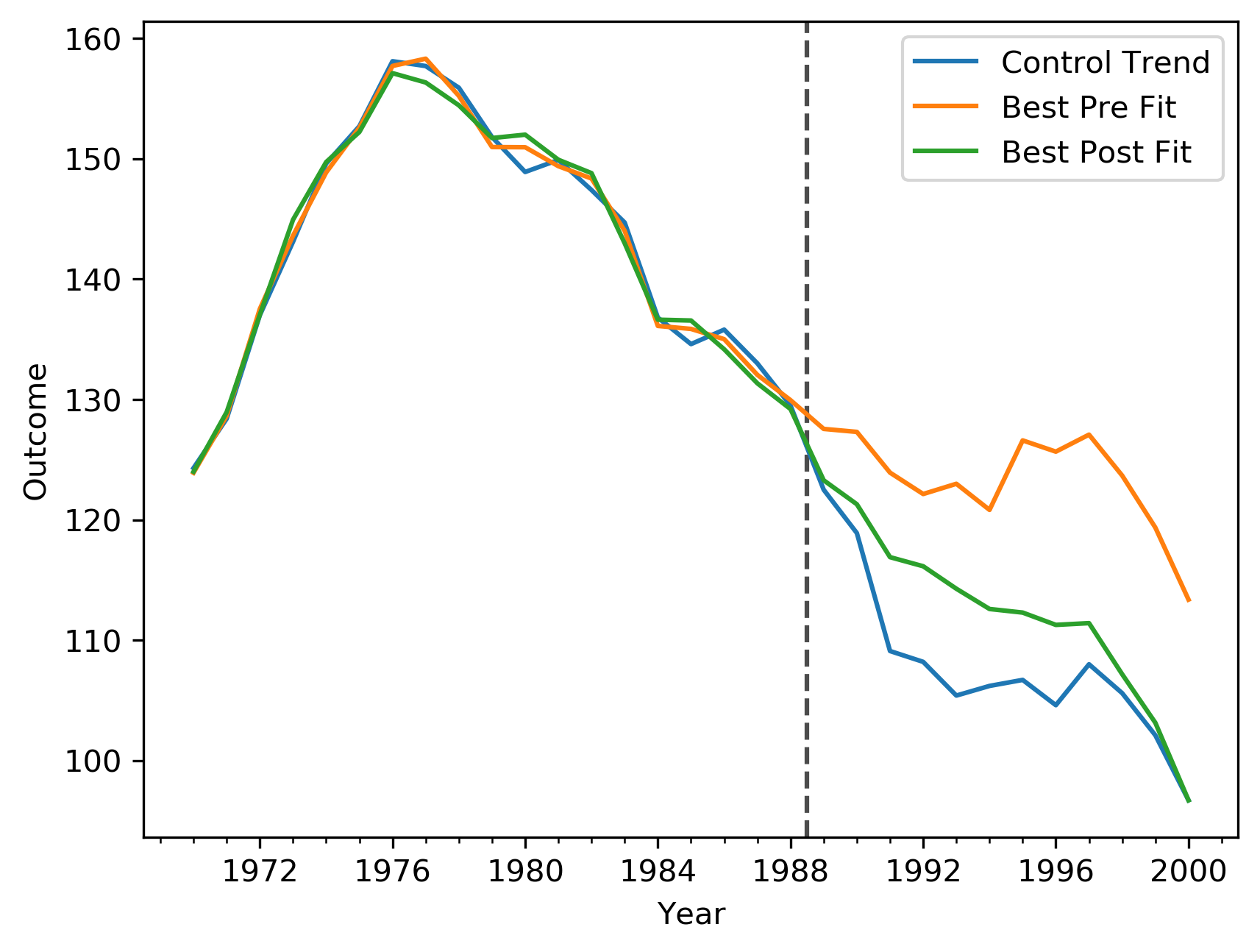}
	\caption{Virginia Placebo Analysis}
	\label{fig:intro_virginia}
	\end{subfigure}
	\begin{subfigure}{0.5\textwidth}
	\centering
	\includegraphics[width=\textwidth]{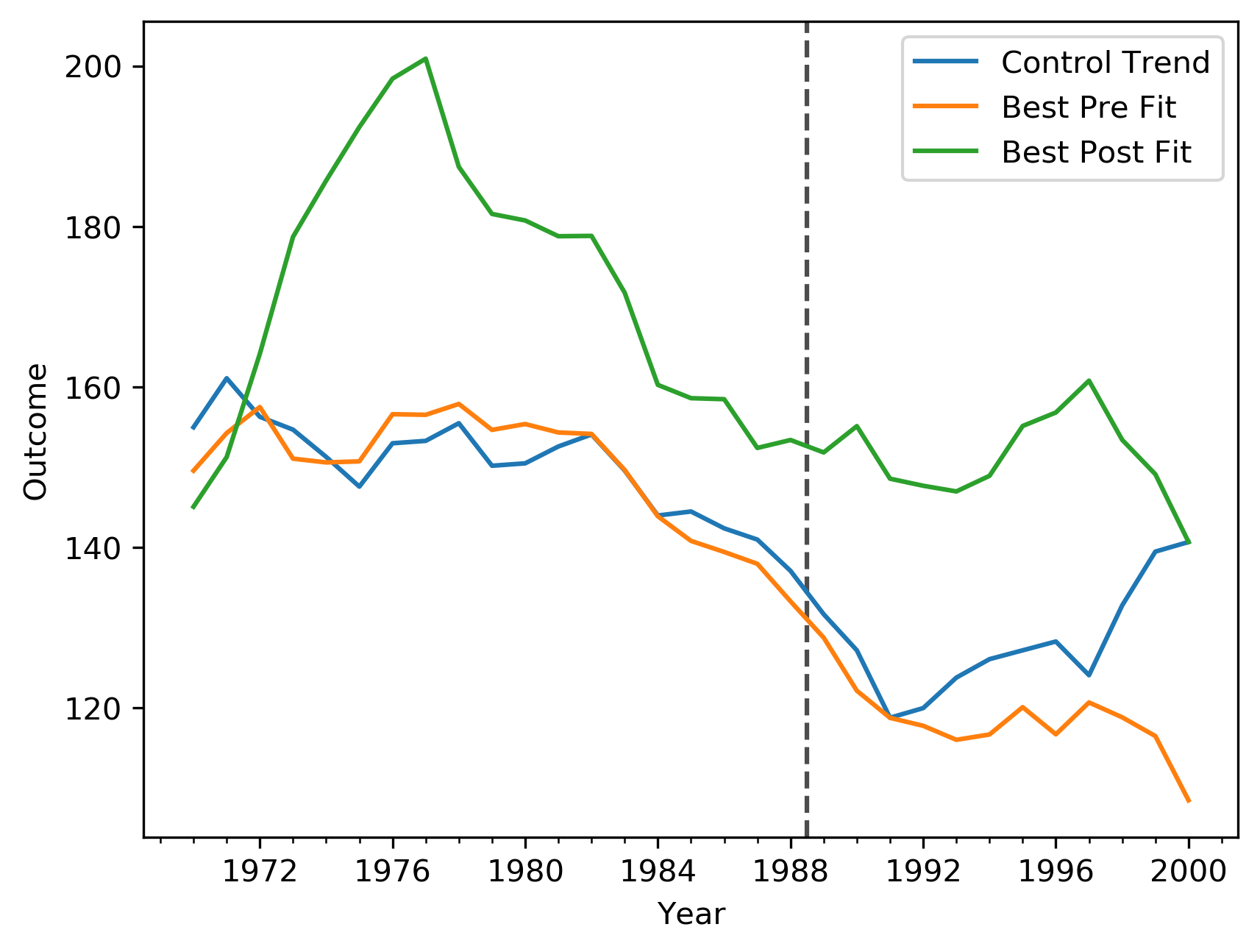}
	\caption{Delaware Placebo Analysis}
	\label{fig:intro_delaware}
	\end{subfigure}
	\caption{Visualizations of placebo analyses with Virginia and Delaware as treated units; in both panels, the blue trend denotes the true control outcome trend of the placebo treated unit, the orange trend denotes the synthetic control trend selected by the SC method, and the green trend denotes the ``best-looking'' synthetic control trend that minimizes pre-treatment fit error while exactly matching the placebo treated unit's control outcome in 2000.}
	\label{fig:intro_placebo}
	\end{figure}

In Figure \ref{fig:intro_virginia}, we depict the observed, true control outcomes for Virgina alongside two different convex combinations of the remaining control units' outcome trends. The first is the orange synthetic control trend constructed in typical SC fashion, namely as the convex combination of control units' outcomes that most closely approximates Virginia's pre-treatment outcomes \citep{abadie2010synthetic}. While this procedure yields a trend with good pre-treatment fit, it does a subpar job of predicting Virginia's control outcome in the year 2000. 

Next, since we observe Virginia's control outcomes post-treatment in this placebo analysis, we can instead construct the ``best-looking'' (in a pre-treatment fit sense) convex combination of the remaining control units' outcome trends that matches Virginia's control outcome in 2000 exactly, shown in green.\footnote{We will discuss how we can compute such a convex combination in Section \ref{subsec:misspec_generalize}. Note that doing so is only possible because Virginia's control outcome in 2000 lies between the minimum and maximum of the other control units' outcomes in 2000, in which case there are many convex combinations with no prediction error.} Perhaps surprisingly, there exists a convex combination of control units that exactly predicts our post-treatment outcome of interest while achieving only marginally worse pre-treatment fit than the best-fitting trend chosen by the SC method.

When we conduct the same exercise with Delaware as the ``treated'' unit of interest, we see in Figure \ref{fig:intro_delaware} that, just as with Virginia, although the trend constructed by the SC method (again in orange) has good-looking pre-treatment fit, it does a poor job estimating the true control outcome of interest. However, unlike when we used Virginia as the placebo treated unit, we cannot construct a convex combination of control units' outcome trends that matches both Delaware's control outcome in 2000 exactly and its pre-treatment outcomes well, so the best-looking convex combination of control units' trends we select to match Delaware's outcome in 2000 exactly (again in green) has unacceptable pre-treatment fit.

These two examples indicate we should interpret SC estimates with caution; the placebo analysis using Virginia suggests good pre-treatment fit is not sufficient for good post-treatment accuracy, while the placebo analysis using Delaware suggests good pre-treatment accuracy, while feasible, may not even be achievable alongside post-treatment accuracy. As discussed in Section \ref{subsec:misspec_generalize}, the additional pre-treatment fit error incurred by the green trends beyond the minimum error incurred by the orange trends is one natural measure of \emph{misspecification error}.

This perspective on the informativeness of pre-treatment fit (or lack thereof) is also the motivation for our proposed sensitivity analysis. While we do not observe the treated unit's counterfactual post-treatment outcomes and thus cannot compute its misspecification error, we can compute the misspecification errors incurred by the SC method when we use it to predict control units' post-treatment outcomes, as in the placebo analyses of Virginia and Delaware. Our procedure assumes the treated unit's misspecification error is at most the misspecification error of a given control unit and computes the set of treatment effects consistent with the assumption that the unknown misspecification error incurred by the SC method is at most this error bound. 

\begin{figure}[t]
	\captionsetup{width=0.93\textwidth}
	\begin{subfigure}{0.5\textwidth}
	\centering
	\includegraphics[width=\textwidth]{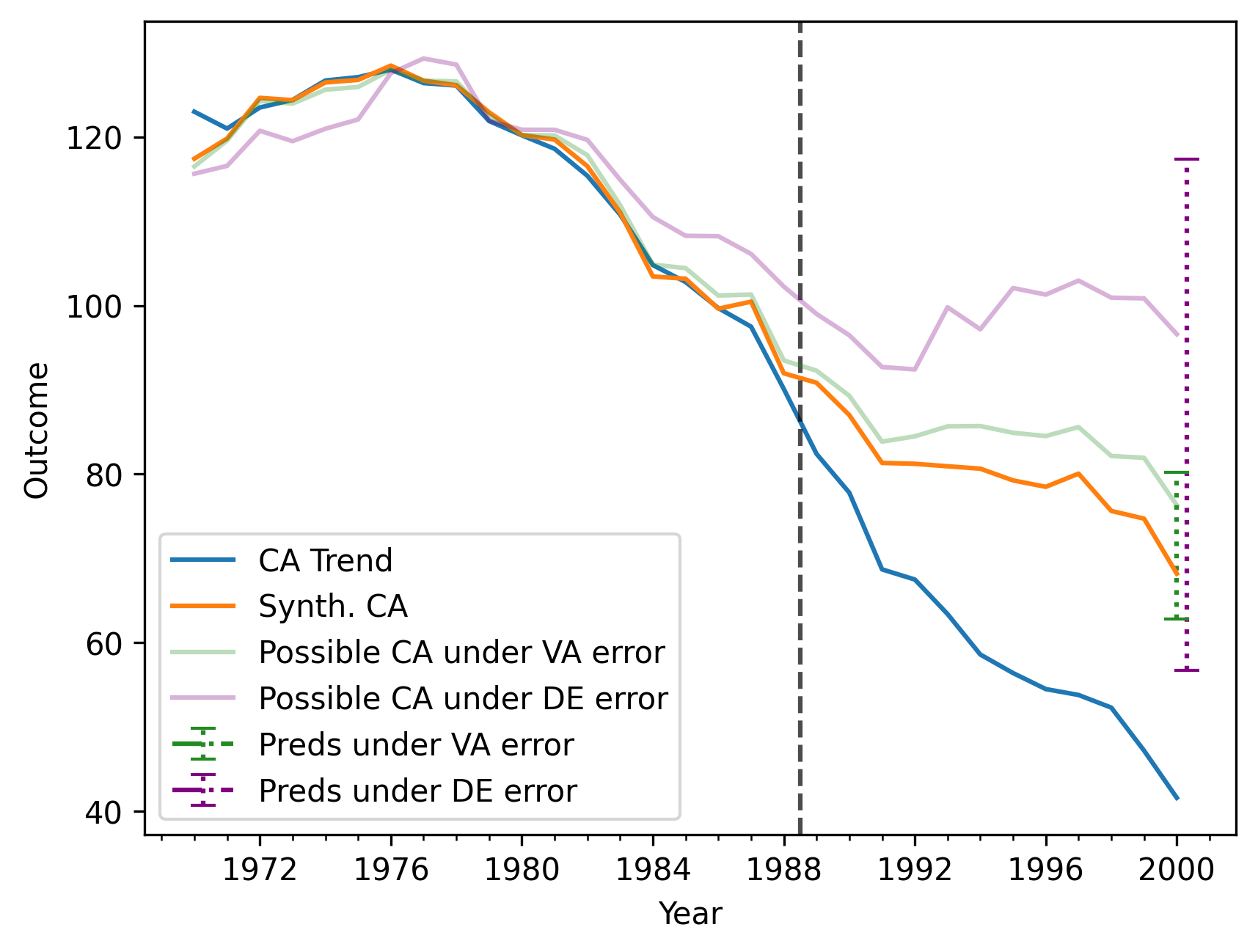}
	\caption{California SC Predictions}
	\label{fig:intro_california}
	\end{subfigure}
	\begin{subfigure}{0.5\textwidth}
	\centering
	\includegraphics[width=\textwidth]{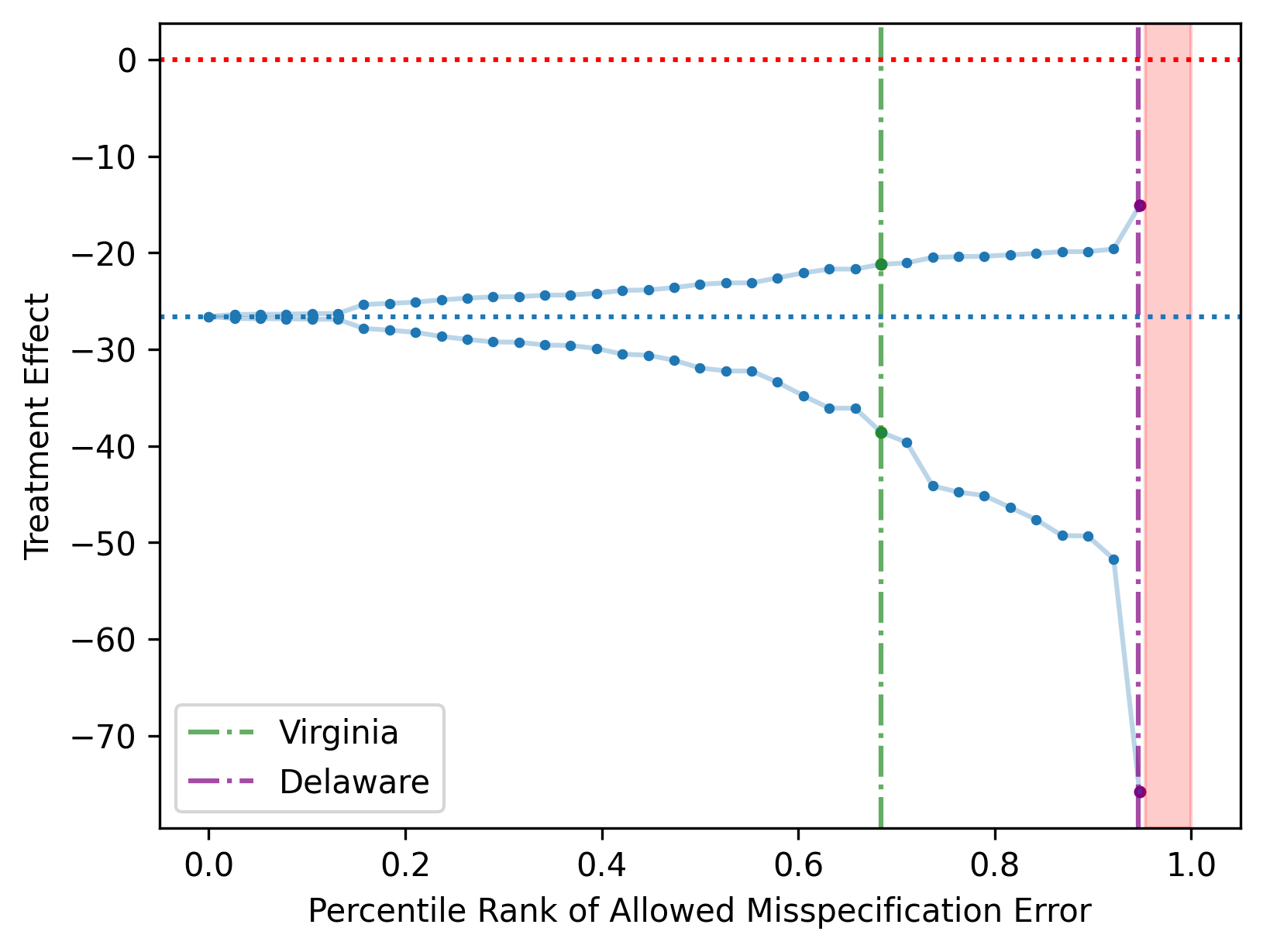}
	\caption{California Treatment Effect Bounds}
	\label{fig:intro_bounds_plot}
	\end{subfigure}
	\caption{In Figure \ref{fig:intro_california}, the blue trend denotes the observed outcome trend for California and the orange trend denotes the synthetic control trend selected by the SC method. The green and purple dotted intervals depict the sets of plausible counterfactual control outcomes for California under Virginia and Delaware's misspecification errors respectively, and we provide  examples of counterfactual trends for California that satisfy the error bound for Virginia in light green and the error bound for Delaware in light purple. Figure \ref{fig:intro_bounds_plot} presents the plausible treatment effect bounds corresponding to the misspecification errors of all the control units in the donor pool in order of error magnitude, with the bounds for Virginia and Delaware highlighted with green and purple dashed lines. The red region indicates where in the distribution of misspecification errors a zero treatment effect first becomes plausible.}
	\label{fig:intro_treated}
	\end{figure}

In Figure \ref{fig:intro_california}, we depict the sets of plausible counterfactual control outcomes for California computed by our procedure consistent with the assumptions that the SC method's misspecification error for California is at most the observed misspecification errors for Virginia and Delaware, indicated by the green and purple dotted intervals respectively. For intuition, we also include examples of predicted counterfactual trends for California that satisfy these error bounds in light green for Virginia and light purple for Delaware. 

Our procedure also finds the minimum misspecification error necessary for zero to be a plausible treatment effect, which we can use to assess how reasonable a zero treatment effect would be by benchmarking it against the placebo misspecification errors described above. In Figure \ref{fig:intro_bounds_plot}, we show the treatment effect bounds corresponding to each control unit's misspecification error in order of increasing error magnitude, and we use the red region to highlight where in the distribution of misspecification errors a zero treatment effect first becomes plausible.\footnote{As we discuss in more detail in Section \ref{subsec:gen_sensitivity_analysis}, the control units with the largest and smallest post-treatment outcomes in 2000 cannot be perfectly predicted using a convex combination of the remaining control units' outcomes, so the misspecification errors incurred by the SC method for these two placebo treated units are infinite. As a result, the sets of plausible treatment effects corresponding to these extremal units mechanically span the real line, so they cannot be plotted, and they cannot rule out a zero treatment effect.} We also illustrate how Virginia and Delaware's misspecification errors compare to those of the other control units by highlighting the treatment effect bounds corresponding to Virginia and Delaware's misspecification errors with green and purple dashed lines.

Although in general, our proposed treatment effect bounds must be computed numerically using convex programming tools \citep{boyd2004convex}, applying our procedure with misspecification error defined as the minimum distance between the SC weights and any vector of weights with perfect predictive accuracy yields closed-form bounds whose widths are determined by scaled-up residuals from predicting control units' outcomes with the SC method. As such, one can view our procedure applied with this misspecification error metric as a geometric motivation for a more conservative variant of the popular randomization inference-based placebo test of no treatment effect proposed in \cite{abadie2010synthetic}. When we apply our procedure to several canonical studies that report SC estimates, we broadly confirm the conclusions drawn by the source papers. We also demonstrate via placebo analyses using these datasets that, in contrast with our proposed procedure, popular robustness checks for SC estimates are plagued by ambiguities in implementation and interpretation and do not fully characterize the extent to which misspecification error can affect the validity of SC estimates.

Importantly, our analysis assumes that errors in SC estimates are driven by model misspecification, not statistical noise, since often, it is not clear what stochastic data generating processes are appropriate models of comparative case study settings with small donor pools containing heterogeneous units observed over short time horizons and selected in a potentially non-random fashion \citep{abadie2020using}. Such an approach is not unprecedented; given ambiguity about the appropriateness of various sampling frameworks in comparative case study settings, \cite{manski2018right} do not specify a sampling model and focus instead on assessing estimate sensitivity to modeling assumptions, which they argue is a crucial and often-overlooked source of uncertainty. Moreover, applied researchers are already accustomed to using non-statistical procedures to assess SC estimate credibility like the popular robustness checks we discuss in Section \ref{subsec:other_robustness}. In line with this perspective, our analysis should not be interpreted as a statistical inference procedure, but rather as a complement to existing statistical approaches for assessing uncertainty in SC estimates like those proposed in \cite{abadie2010synthetic}, \cite{firpo2018synthetic}, \cite{chernozhukov2017exact}, \cite{chernozhukov2018practical}, \cite{cattaneo2019prediction}, and \cite{li2019statistical}.

The rest of the paper proceeds as follows. Section \ref{sec:sensitivty_analysis} introduces a version of our proposed procedure that admits a particularly simple form, as summarized in Procedure \ref{alg:weight_space_sensitivity_analysis} and applied in Section \ref{sec:case_studies} to several canonical studies that use the SC method. In the course of introducing our method, we also discuss how it is related to a placebo test suggested in \cite{abadie2010synthetic}, how donor pool selection impacts both our procedure and SC treatment effect estimates more broadly, and how our approach compares to existing robustness checks popular in the SC literature. Then, in Section \ref{subsec:misspec_generalize}, we provide a general sensitivity analysis framework outlined in Procedure \ref{alg:general_sensitivity_analysis} that can accommodate other valuable notions of misspecification error, which we discuss in the context of the studies revisited earlier.

\section{Sensitivity Analysis}\label{sec:sensitivty_analysis}

To introduce the ideas underlying our sensitivity analysis, in this section, we present a particular version of our proposed procedure based on a measure of misspecification error that yields intuitive, closed-form expressions for our treatment effect bounds. Later, we generalize the procedure to accommodate other valuable notions of misspecification error.

\subsection{Notation and the SC Method}\label{subsec:notation}

Before describing our procedure, we introduce some necessary notation and review the SC method. In the canonical setting used to motivate the SC method, we acquire data about $J + 1$ units across $T$ time periods $[T] \coloneqq \{1, \dots, T\}$ to estimate the effect of a policy intervention, referred to as the treatment, affecting a single treated unit indexed by $j = 1$. The treatment is first implemented just after $T_0 < T$ and stays in effect for all remaining periods $T_0 + 1, \dotsc, T$. The set of $J$ remaining control units $\mathcal{J} \coloneqq \{2, \dots, J+1\}$ that are not affected by the treatment is called the \emph{donor pool}. For each unit $j \in \{1, \dots, J+1\}$ and each time period $t \in [T]$, we let $Y_{jt}(1)$ and $Y_{jt}(0)$ denote that unit's potential outcomes in that period under treatment and lack thereof, respectively \citep{imbens2015causal}. Next, let the indicator $D_{jt} = 1$ if unit $j$ is exposed to the treatment in period $t$ and $D_{jt} = 0$ otherwise. We then let $Y_{jt} \coloneqq D_{jt}Y_{jt}(1) + (1-D_{jt})Y_{jt}(0)$ denote the potential outcome we observe for unit $j$ in period $t$, and let $\mathbf{Y}_{0t} \coloneqq \left(Y_{2t}, \dotsc, Y_{(J+1)t}\right)^T$ denote the vector of control units' observed outcomes in period $t$.

Typically, the goal in comparative case study settings like these is to estimate the treatment effect on the treated unit (with index $j = 1$) in some post-treatment period $T^* > T_0$:
\[
\tau_{T^*} \coloneqq Y_{1T^*}(1) - Y_{1T^*}(0).
\]
Because we only observe $Y_{1T^*} = Y_{1T^*}(1)$, and not $Y_{1T^*}(0)$, estimating $\tau$ reduces to estimating $Y_{1T^*}(0)$. Although there are many ways one could do so, the SC method assumes it is possible to compute $Y_{1T^*}(0)$ using a weighted sum of the control units' outcomes in period $T^*$ \citep{abadie2003basque,abadie2010synthetic}:
\begin{equation}\label{eq:well-specification}
\begin{aligned}
Y_{1T^*}(0) &= \mathbf{Y}_{0T^*}^T\mathbf{w} = \sum_{j = 2}^{J+1}w_jY_{jT^*}(0) \quad \text{for $\mathbf{w} = (w_2, \dotsc, w_{J+1})^T \in \R^J$}.
\end{aligned}
\end{equation}
We refer to this weighted combination of control units' outcome trends as a \emph{synthetic control}.

In particular, \cite{abadie2010synthetic} propose choosing weights $\mathbf{w}$ that make the weighted average of the control units' pre-treatment outcomes as similar as possible to the treated unit's pre-treatment outcomes.\footnote{\cite{abadie2010synthetic}, \cite{chernozhukov2017exact}, and \cite{ferman2019synthetic} discuss several models under which such an assumption is reasonable.} Let $\mathbf{x}_{j} \coloneqq \left(Y_{j1}, \dotsc, Y_{jT_0}\right)^T$ be the vector of unit $j$'s observed, pre-treatment outcomes and $X_0$ be the $T_0 \times J$ matrix whose columns are the control units' observed pre-treatment outcomes (i.e. $X_0$'s $j$th column is given by $\mathbf{x}_{j+1}$). Then we can write the SC estimator as the minimizer of pre-treatment prediction error over the set of positive weights that sum to one:\footnote{While uncommon in practice, $\mathbf{x}_1$ could in principle lie in the convex hull of the columns of $X_0$, in which case \eqref{eq:basic_synth_contr_method} could have an infinite number of solutions with perfect pre-treatment fit, some of which would provide better post-treatment fit than others \citep{abadie2020using}. In the sections that follow, we assume perfect pre-treatment fit is not achievable because it is empirically rare and doing so allows us to develop the more intuitive sensitivity analysis presented in Section \ref{sec:sensitivty_analysis}. However, in Section \ref{sec:mult_solns_appendix} of the Appendix, we discuss in detail how the generalized sensitivity analysis described in Section \ref{subsec:gen_sensitivity_analysis} can easily account for non-uniqueness of the SC estimator when generating treatment effect bounds. We also note that \cite{crest2018penalized} and \cite{kellogg2020combining} propose modifying the SC objective to penalize solutions that interpolate more between units, since such solutions will yield worse predictions if the relationship between pre-treatment outcomes and post-treatment outcomes is nonlinear. Our sensitivity analysis can also be applied to these alternative estimators, as we detail in Section \ref{subsec:lpes} of the Appendix. \label{foot:mult_solns}}

\begin{equation}\label{eq:basic_synth_contr_method}
\begin{aligned}
\mathbf{w}_\text{sc} \coloneqq \argmin_{\mathbf{w} \in \R^J}&~ \norm{\mathbf{x}_{1} - X_{0}\mathbf{w}}_2 \\
\suchthat&~ \ones^T\mathbf{w} = 1  \\ 
&~ \mathbf{w} \geq \zeros 
\end{aligned}
\end{equation}
Later, we will use $\Delta_J \coloneqq \left\{\mathbf{w} \in \R^J \setst \mathbf{w} \geq \zeros, ~\ones^T\mathbf{w} = 1\right\}$ to denote the set of valid SC weights.

Once $\mathbf{w}_\text{sc}$ has been computed, we can estimate $\tau_{T^*}$ by
\[
\hat{\tau}^\text{sc}_{T^*} \coloneqq Y_{1T^*} - \mathbf{Y}_{0T^*}^T\mathbf{w}_\text{sc} = Y_{1T^*}(1) - \sum_{j = 2}^{J+1}w_{\text{sc},j}Y_{jT^*}(0).
\]
If \eqref{eq:well-specification} holds for weights $\mathbf{w} = \mathbf{w}_\text{sc}$, we say the SC method is \emph{well-specified}, in which case we have that $\hat{\tau}^\text{sc}_{T^*} = \tau_{T^*}$. However, as illustrated in Section \ref{sec:intro}, the SC method is unlikely to be well-specified in practice. In the next section, we will introduce a natural way to measure the degree to which the SC method deviates from well-specification on which we will base our sensitivity analysis.

In Section \ref{subsec:lpes} of the Appendix, we discuss how the sensitivity analyses we introduce below can also apply to extensions of the SC method that incorporate additional pre-treatment covariates, relax the convex weight constraints, add an intercept term, and minimize different and sometimes data-adaptive objective functions. For expositional clarity however, the basic SC method presented here will suffice to motivate our proposed procedures.

\subsection{The Procedure}

\subsubsection{Bounding Treatment Effects Under Misspecification}\label{subsec:bounding_tes_misspec}

Despite the concerns about the effectiveness of the SC method raised in Section \ref{sec:intro}, we can still attempt to assess what the true value of $Y_{1T^*}(0)$ might be under limited misspecification error. Since $\hat{\tau}^\text{sc}_{T^*} = Y_{1T^*}(1) - \mathbf{Y}_{0T^*}^T\mathbf{w}_\text{sc}$ is an affine function of $\mathbf{Y}_{0T^*}$ and $\tau_{T^*}$ is a scalar, it is always possible to choose some set of weights $\mathbf{w} \in \R^J$ such that $\mathbf{Y}_{0T^*}^T\mathbf{w} = Y_{1T^*}(0)$ and thus $Y_{1T^*}(1) - \mathbf{Y}_{0T^*}^T\mathbf{w} = \tau_{T^*}$. More importantly, they are not at all unique; in fact, the set of optimal weights $$\mathcal{W}_1^* \coloneqq \{\mathbf{w} \in \R^J \setst \mathbf{Y}_{0T^*}^T\mathbf{w} = Y_{1T^*}(0)\}$$ forms a $(J-1)$-dimensional hyperplane in $\R^J$. Thus, a natural measure of misspecification error in the SC weights $\mathbf{w}_\text{sc}$ is the difference between $\mathbf{w}_\text{sc}$ and the closest weights $\mathbf{w}_*$ to $\mathbf{w}_\text{sc}$ in $\mathcal{W}_1^*$, where distance is measured by the $\ell_2$-norm. More formally, we can define $\mathbf{w}_*$ like so:
\begin{equation}\label{eq:min_dist_to_opt_weights_treated_unit}
\begin{aligned}
\mathbf{w}_* \coloneqq \argmin_{\mathbf{w} \in \R^J}&~ \norm{\mathbf{w}_\text{sc} - \mathbf{w}}_2 \\
\suchthat&~ \mathbf{Y}_{0T^*}^T\mathbf{w} = Y_{1T^*}(0) ~~\left(\Leftrightarrow \mathbf{w} \in \mathcal{W}_1^*\right).
\end{aligned}
\end{equation}
Note that we do not restrict ourselves to considering weights within the set of convex weights $\Delta_J$; though such a restriction prevents SC estimates from extrapolating beyond the outcomes in the data \citep{abadie2020using}, it may be that the closest weights that allow for optimal prediction of $Y_{1T^*}(0)$ lie outside $\Delta_J$, or that $\mathcal{W}_1^*$ and $\Delta_J$ do not overlap at all. In what follows, we will frequently focus on the magnitude of misspecification error, which we denote by $d_2\left(\mathbf{w}_\text{sc}, \mathcal{W}_1^*\right) \coloneqq \norm{\mathbf{w}_\text{sc} - \mathbf{w}_*}_2$.

Since $\mathcal{W}_1^*$ is a hyperplane, we could in principle solve \eqref{eq:min_dist_to_opt_weights_treated_unit} by projecting $\mathbf{w}_\text{sc}$ onto $\mathcal{W}_1^*$. Because we do not observe $Y_{1T^*}(0)$, we cannot do so in practice. However, if we are willing to assume $d_2\left(\mathbf{w}_\text{sc}, \mathcal{W}_1^*\right) \leq B$ for some bound $B \geq 0$, then there must be some weight vector $\mathbf{w} \in \R^J$ within a radius $B$ $\ell_2$-ball around $\mathbf{w}_\text{sc}$ such that $\mathbf{Y}_{0T^*}^T\mathbf{w} = Y_{1T^*}(0)$. Crucially, this assumption limits the magnitude of method misspecification error while allowing for the direction of that error to remain arbitrary. If we let $$\widehat{\mathcal{W}}_1^B \coloneqq \left\{\mathbf{w} \in \R^J \setst \norm{\mathbf{w}_\text{sc} - \mathbf{w}}_2 \leq B\right\}$$ denote the set of all weights $\ell_2$-distance at most $B$ away from $\mathbf{w}_\text{sc}$, then we know that the true potential outcome $Y_{1T^*}(0)$ lies within the following set of values: $$\mathcal{Y}_{1T^*}^B(0) \coloneqq \mathbf{Y}_{0T^*}^T\widehat{\mathcal{W}}_1^B = \left\{\mathbf{Y}_{0T^*}^T\mathbf{w} \setst \norm{\mathbf{w}_\text{sc} - \mathbf{w}}_2 \leq B\right\}.$$

Since the function $\mathbf{w} \mapsto \mathbf{Y}_{0T^*}^T\mathbf{w}$ is continuous in $\mathbf{w}$ and $\widehat{\mathcal{W}}_1^B$ is compact, the set $\mathcal{Y}_{1T^*}^B(0)$ containing $Y_{1T^*}(0)$ must be a closed interval in $\R$. As a result, we can characterize the interval $\mathcal{Y}_{1T^*}^B(0)$ by computing its endpoints $Y_{1T^*}^{B,-}(0)$ and $Y_{1T^*}^{B,+}(0)$, which are the solutions to the following two optimization problems:
\begin{equation}\label{eq:bounds_on_cntrfct_control_outcomes}
\begin{aligned}
Y_{1T^*}^{B,-}(0) \coloneqq \min_{\mathbf{w} \in \R^J} &~ \mathbf{Y}_{0T^*}^T\mathbf{w} \quad& Y_{1T^*}^{B,+}(0) \coloneqq \max_{\mathbf{w} \in \R^J} &~ \mathbf{Y}_{0T^*}^T\mathbf{w} \\
\suchthat&~ \norm{\mathbf{w}_\text{sc} - \mathbf{w}}_2 \leq B \quad& \suchthat&~ \norm{\mathbf{w}_\text{sc} - \mathbf{w}}_2 \leq B
\end{aligned}
\end{equation}
Since $Y_{1T^*}^{B,-}(0)$ and $Y_{1T^*}^{B,+}(0)$ are defined as the extrema of linear functions on an $\ell_2$-ball centered at $\mathbf{w}_\text{sc}$, they can easily be computed in closed form:
\begin{equation*}
\begin{aligned}
	Y_{1T^*}^{B,-}(0) &= \mathbf{Y}_{0T^*}^T\mathbf{w}_\text{sc} - B\normnofit{\mathbf{Y}_{0T^*}}_2 \quad& Y_{1T^*}^{B,+}(0) &= \mathbf{Y}_{0T^*}^T\mathbf{w}_\text{sc} + B\normnofit{\mathbf{Y}_{0T^*}}_2.
\end{aligned}
\end{equation*}
Then, since $\tau_{T^*}$ is linear in $Y_{1T^*}(0)$, we can translate these bounds on $Y_{1T^*}(0)$ into bounds on $\tau_{T^*}$:
\begin{equation}\label{eq:bounds_on_treat_eff_gen_B}
\begin{aligned}
	\tau_{T^*} \in \mathcal{T}^B_{T^*} &\coloneqq \left[Y_{1T^*}(1) - Y_{1T^*}^{B,+}(0), Y_{1T^*}(1) - Y_{1T^*}^{B, -}(0)\right] \\
	&= \hat{\tau}^\text{sc}_{T^*} + B\normnofit{\mathbf{Y}_{0T^*}}_2 \cdot [-1, 1] 
\end{aligned}
\end{equation}

\subsubsection{Bound Calibration via Placebo Effect Estimation}\label{sec:bound_calibration}

Unfortunately, the discussion in Section \ref{subsec:bounding_tes_misspec} does not make it clear how one should choose an appropriate misspecification error bound $B$ from which the bounds on $\tau_{T^*}$ in \eqref{eq:bounds_on_treat_eff_gen_B} can be constructed. However, since we do observe $Y_{jT^*} = Y_{jT^*}(0)$ for each control unit $j \in \mathcal{J}$, we can use a similar distance measure to $d_2\left(\mathbf{w}_\text{sc}, \mathcal{W}^*_1\right)$ to quantify the misspecification error in SC estimates of $Y_{jT^*}(0)$ for $j \in \mathcal{J}$ using the remaining $J - 1$ control units as donor pools. Then, we can assume the treated unit's post-treatment potential outcome $Y_{1T^*}(0)$ is no more difficult to estimate using the SC method than some percentage of the control units' post-treatment control outcomes and use these measures to inform our choice of bound $B$.

Importantly, the methodology we propose below based on this intuition only relies on the assumption that the magnitude of the misspecification error for the treated unit is no larger than the magnitudes of the placebo misspecification errors for some percentage of the control units. Given that the differences in characteristics between the treated and control units is a primary reason researchers should use the SC method in the first place \citep{abadie2020using}, it is likely implausible that the unknown direction of the treated unit's misspecification error is similar to the directions of the control units' placebo misspecification errors.

To formalize the ideas presented above, we first define the following quantities analogous to $X_0$, $\mathbf{Y}_{0T^*}$, $\mathbf{w}_\text{sc}$, and $\mathcal{W}^*_1$ when we view control unit $j$ as a placebo treated unit and the other $J-1$ control units as the donor pool: let $X_{-j}$ be the $T_0 \times (J-1)$ matrix whose columns are the pre-treatment outcomes of the $J-1$ control units other than $j$ \footnote{i.e. $X_{-j}$'s $k$th column is given by $\mathbf{x}_{k+1}$ if $k < j$ and $\mathbf{x}_{k+2}$ if $k > j$.}, let $\mathbf{Y}_{(-j)T^*} \coloneqq \left(Y_{kt}\right)_{k \neq j}^T$ be the $(J-1)$-vector of the $J-1$ control units besides $j$'s observed control outcomes, let $\mathbf{w}_\text{sc}^{(j)} \in \R^{J-1}$ be the synthetic control weights chosen as if control unit $j$ were the treated unit and the remaining $J-1$ control units were the donor pool, i.e. by solving the following optimization problem similar to \eqref{eq:basic_synth_contr_method}:
\begin{equation}\label{eq:basic_synth_contr_method_placebo}
\begin{aligned}
\mathbf{w}_\text{sc}^{(j)} \coloneqq \argmin_{\mathbf{w} \in \R^{J-1}}&~ \norm{\mathbf{x}_{j} - X_{-j}\mathbf{w}}_2 \\
\suchthat&~ \ones^T\mathbf{w} = 1 \\
&~ \mathbf{w} \geq \zeros,
\end{aligned}
\end{equation}
and let $\mathcal{W}^*_j \coloneqq \{\mathbf{w} \in \R^{J-1} \setst \mathbf{Y}_{(-j)T^*}^T\mathbf{w} = Y_{jT^*}(0)\}$ denote the set of weight vectors $\mathbf{w} \in \R^{J-1}$ that yield placebo unit $j$'s control outcome in period $T^*$.

Since we observe $Y_{jT^*} = Y_{jT^*}(0)$ for placebo unit $j$, we can actually compute the distance $d_2(\mathbf{w}_\text{sc}^{(j)}, \mathcal{W}^*_j)$ defined analogously to the unobservable $d_2\left(\mathbf{w}_\text{sc}, \mathcal{W}^*_1\right)$ in \eqref{eq:min_dist_to_opt_weights_treated_unit}:
\begin{equation}\label{eq:min_dist_to_opt_weights_control_unit}
	\begin{aligned}
		d_2(\mathbf{w}_\text{sc}^{(j)}, \mathcal{W}^*_j) \coloneqq \min_{\mathbf{w} \in \R^{J-1}}&~ \normnofit{\mathbf{w}_\text{sc}^{(j)} - \mathbf{w}}_2 \\
	\suchthat&~ \mathbf{Y}_{(-j)T^*}^T\mathbf{w} = Y_{jT^*} ~~\left(\Leftrightarrow \mathbf{w} \in \mathcal{W}_j^*\right).
	\end{aligned}
	\end{equation}
For notational convenience, let $\hat{R}_{jT^*}^\text{sc} \coloneqq \mathbf{Y}_{(-j)T^*}^T\mathbf{w}^{(j)}_\text{sc} - Y_{jT^*}$ denote the residual from the SC estimator used to predict $Y_{jT^*} = Y_{jT^*}(0)$. Then as with \eqref{eq:min_dist_to_opt_weights_treated_unit}, \eqref{eq:min_dist_to_opt_weights_control_unit} is a basic projection problem with a closed-form solution (see \cite{cheney2009linear}, pages 450--451, for example):
\begin{equation}\label{eq:def_dist_to_opt}
	d_2(\mathbf{w}_\text{sc}^{(j)}, \mathcal{W}_j^*) = \frac{\abs{\hat{R}_{jT^*}^\text{sc}}}{\norm{\mathbf{Y}_{(-j)T^*}}_2} = \frac{\absfit{\mathbf{Y}_{(-j)T^*}^T\mathbf{w}_\text{sc}^{(j)} - Y_{jT^*}}}{\norm{\mathbf{Y}_{(-j)T^*}}_2}
\end{equation}

Although $d_2(\mathbf{w}_\text{sc}^{(j)}, \mathcal{W}_j^*)$ is defined purely geometrically, choosing the $\ell_2$-norm to measure distance in weight space implies $d_2(\mathbf{w}_\text{sc}^{(j)}, \mathcal{W}_j^*)$ can also be characterized as a scaled variant of the absolute placebo SC residual $\abs{\hat{R}_{jT^*}^\text{sc}}$ for control unit $j$ using the $j-1$ other control units as the donor pool. We will discuss this observation in more detail in Section \ref{sec:interpretation}.

For notational convenience, we use the shorthand $B_j = d_2(\mathbf{w}_\text{sc}^{(j)}, \mathcal{W}^*_j)$ and assume control units' indices align with the sorted order of their respective $B_j$ values, so that the $j$th control unit has the $(j-1)$th-smallest $B_j$. Then once we have computed $d_2(\mathbf{w}_\text{sc}^{(j)}, \mathcal{W}^*_j)$ for all $j \in \mathcal{J}$, we can compute bounds on the treatment effect based on \eqref{eq:bounds_on_treat_eff_gen_B} for each $j \in \mathcal{J}$ by choosing $B = B_j \coloneqq d_2(\mathbf{w}_\text{sc}^{(j)}, \mathcal{W}^*_j)$:
\begin{equation}\label{eq:def_T_B_j_bounds}
	\mathcal{T}^{B_j}_{T^*} = \hat{\tau}_{T^*} + \abs{\hat{R}_{jT^*}^\text{sc}}\frac{\normnofit{\mathbf{Y}_{0T^*}}_2}{\normnofit{\mathbf{Y}_{(-j)T^*}}_2} \cdot [-1, 1]
\end{equation}

Another natural quantity of interest is the minimum bound $B_0$ on $d_2\left(\mathbf{w}_\text{sc}, \mathcal{W}^*_1\right)$ such that a zero treatment effect lies within $\mathcal{T}_{T^*}^{B_0}$, i.e. $B_0 \coloneqq \min \left\{B \setst 0 \in \mathcal{T}_{T^*}^B\right\}$. With $B_0$ in hand, we can then find the control unit $j_0 \in \mathcal{J}$ such that $B_{j_0} \leq B_0 \leq B_{j_0+1}$ (where $B_{J+2} = \infty$) and report the statistic $\nu \coloneqq (j_0-1)/J$, interpreted as the fraction of control units for which it would have to be ``easier'' for the SC method to estimate $Y_{jT^*}(0)$ than $Y_{1T^*}(0)$ if the treatment effect $\tau_{T^*}$ for the treated unit were actually zero. For the purposes of computation, $B_0$ can be defined similarly to $d_2\left(\mathbf{w}_\text{sc}, \mathcal{W}_1^*\right)$, with the unobserved $Y_{1T^*}(0)$ replaced with the observed outcome $Y_{1T^*}$ of the treated unit in period $T^*$:
\begin{equation}\label{eq:nullifying_distance_comp_def}
\begin{aligned}
B_0 \coloneqq \min_{\mathbf{w} \in \R^J}&~ \norm{\mathbf{w}_\text{sc} - \mathbf{w}}_2 \\
\suchthat &~ \mathbf{Y}_{0T^*}^T\mathbf{w} = Y_{1T^*}
\end{aligned}
\end{equation}
As with \eqref{eq:min_dist_to_opt_weights_control_unit}, $B_0$ can easily be computed in closed form by projecting $\mathbf{w}_\text{sc}$ onto the hyperplane $\left\{\mathbf{w} \in \R^J \setst \mathbf{Y}_{0T^*}^T\mathbf{w} = Y_{1T^*}\right\}$:
\begin{equation}\label{eq:min_dist_expression}
	B_0 = \frac{\absfit{\mathbf{Y}_{0T^*}^T\mathbf{w}_\text{sc} - Y_{1T^*}}}{\norm{\mathbf{Y}_{0T^*}}_2}
\end{equation}

For reference, we summarize the sensitivity analysis procedure we have developed above in Procedure \ref{alg:weight_space_sensitivity_analysis}. We also demonstrate one way to visualize $\mathcal{T}_{T^*}^{B_j}$ for each $j \in \mathcal{J}$ along with $B_{j_0}$ and $B_{j_0 + 1}$ in Figure \ref{fig:california_quantile_bounds_plot} using data on California's 1989 tobacco control program analyzed in \cite{abadie2010synthetic}. In the figure, the units of the $x$-axis are percentile ranks $p_j \coloneqq (j-1)/J$ of the ordered set of placebo misspecification errors $\{B_j \setst j \in \mathcal{J}\}$ rather than the units of $B_j$, so that it is easy to read $\nu$ off of the $x$-axis where the red shaded region begins.

\begin{algbox}[t!]
	\myalg{alg:weight_space_sensitivity_analysis}{Sensitivity Analysis}{
	\begin{enumerate}[leftmargin=3ex]
		\item For each control unit $j \in \mathcal{J}$:
		\begin{enumerate}
			\item Use the SC method to predict unit $j$'s outcome in period $T^*$, treating the other $J-1$ control units as the donor pool; compute the observed residual from this prediction $\hat{R}_{jT^*}^\text{sc} = \mathbf{Y}_{(-j)T^*}^T\mathbf{w}^{(j)}_\text{sc} - Y_{jT^*}$. \label{alg:weight_space_placebo_synth_step}
			\item Compute the bounds $\mathcal{T}^{B_j}_{T^*}$ on the treatment effect $\tau_{T^*}$ under the assumption that the misspecification error $d_2(\mathbf{w}_{sc}, \mathcal{W}^*_1)$ incurred by estimating $Y_{1T^*}(0)$ with the SC method is at most the misspecification error $B_j$ incurred by the SC method in Step \ref{alg:weight_space_placebo_synth_step}:
			\begin{equation*}
				\mathcal{T}^{B_j}_{T^*} = \hat{\tau}_{T^*} + \abs{\hat{R}_{jT^*}^\text{sc}}\frac{\normnofit{\mathbf{Y}_{0T^*}}_2}{\normnofit{\mathbf{Y}_{(-j)T^*}}_2} \cdot [-1, 1],
			\end{equation*}
		\end{enumerate}
	
		\item Compute the minimum misspecification error $B_0$ needed for $0 \in \mathcal{T}^{B_0}_{T^*}$, i.e. $0$ to be a plausible treatment effect estimate:
		\begin{equation*}
		B_0 = \frac{\absfit{\mathbf{Y}_{0T^*}^T\mathbf{w}_\text{sc} - Y_{1T^*}}}{\norm{\mathbf{Y}_{0T^*}}_2},
		\end{equation*}
		and find the control unit $j_0$ with the largest misspecification error still smaller than $B_0$, i.e. where $B_{j_0} \leq B_0 \leq B_{j_0+1}$. \label{alg:weight_space_nullifying_error_step}
		
		\item Visualize the treatment effect bounds $\mathcal{T}^{B_j}_{T^*}$ for each $j \in \mathcal{J}$ and the misspecification errors $B_{j_0}$ and $B_{j_0 + 1}$ in a plot like Figure \ref{fig:california_quantile_bounds_plot}, and report the percentage $\nu = (j_0 - 1)/J$ of control units whose misspecification errors $B_j$ are smaller than $B_0$.
	\end{enumerate}
	}
	\end{algbox}

Before proceeding, we make note of several interesting properties of our proposed bounds $\mathcal{T}^{B_j}_{T^*}$. To do so, we define $N_j \coloneqq \normnofit{\mathbf{Y}_{0T^*}}_2 / \normnofit{\mathbf{Y}_{(-j)T^*}}_2$ so we can write $$\mathcal{T}^{B_j}_{T^*} = \hat{\tau}_{T^*} + \abs{\hat{R}_{jT^*}^\text{sc}} N_j \cdot [-1, 1].$$ Since $\mathbf{Y}_{(-j)T^*}$ contains all of the entries of $\mathbf{Y}_{0T^*}$ except $Y_{jT^*}(0)$, we have that $\normnofit{\mathbf{Y}_{(-j)T^*}}_2 \leq \normnofit{\mathbf{Y}_{0T^*}}_2$, so $N_j \geq 1$. Intuitively, this inflation of the placebo residual for unit $j$ in $\mathcal{T}^{B_j}_{T^*}$ corrects for the fact that the placebo SC procedure for estimating $Y_{jT^*}(0)$ has one fewer control unit at its disposal than the SC procedure for estimating $Y_{1T^*}(0)$ and thus has less flexibility to make more extreme predictions than the SC procedure would for our actual task of interest.

Next, we can write $N_j$ as
\begin{equation*}\label{eq:N_j_alt_exp}
	N_j = \sqrt{\frac{\sum_{k = 2}^J Y^2_{kT^*}(0)}{\sum_{k = 2}^J Y^2_{kT^*}(0) - Y^2_{jT^*}(0)}} = \left[1 - \left(\frac{\abs{Y_{jT}(0)}}{\normnofit{\mathbf{Y}_{0T^*}}_2}\right)^2\right]^{-1/2},
\end{equation*}
enabling us to make two more observations. First, $N_j$ is increasing in the magnitude of $Y_{jT^*}(0)$ relative to $\normnofit{\mathbf{Y}_{0T^*}}_2$, meaning $\mathcal{T}^{B_j}_{T^*}$ is wider if unit $j$ has a larger magnitude outcome in period $T^*$ relative to the outcomes of the other control units and thus could generate more extreme predictions if it contributed to the SC predicted outcome.

Second, under mild conditions, the bounds $\mathcal{T}^{B_j}_{T^*}$ converge to purely residual-based bounds $\hat{\tau}_{T^*}^\text{sc} + \abs{\hat{R}_{jT^*}^\text{sc}} \cdot [-1, 1]$ as the size of the donor pool increases. Consider a sequence of donor pools indexed by their sizes, which with an abuse of notation we denote $\left\{\mathcal{J}_J : J \in \N\right\}$. Then, provided that the outcomes of the units in each of the donor pools do not grow too quickly or too slowly in magnitude, i.e. if $$\lim_{J \rightarrow \infty}\max_{j \in \mathcal{J}_J}\frac{\abs{Y_{jT^*}(0)}}{\normnofit{\mathbf{Y}_{0T^*}}_2} \rightarrow 0,$$ the ratios $N_j$ converge uniformly to $1$ as the sample size $J$ increases. As a consequence, for some $\alpha \in [0, 1]$, the bounds $\mathcal{T}^{B_{\ceil{(1-\alpha) J}}}_{T^*}$ calibrated to the $(1-\alpha)$th percentile of the ordered set of placebo distances $B_j$ will shrink towards the bounds $\hat{\tau}_{T^*}^\text{sc} + \abs{\hat{R}_{\ceil{(1-\alpha) J}T^*}^\text{sc}} \cdot [-1, 1]$ as $J \rightarrow \infty$.

\subsection{Interpretation}\label{sec:interpretation}

As described in Section \ref{sec:bound_calibration}, we can view $\mathcal{T}_{T^*}^{B_j}$ as the set of plausible treatment effects for the treated unit if we assume that the magnitude of misspecification error $d_2\left(\mathbf{w}_\text{sc}, \mathcal{W}^*_1\right)$ incurred by estimating $Y_{1T^*}(0)$ with the SC estimator is no larger than the magnitude of misspecification error $d_2(\mathbf{w}_\text{sc}^{(j)}, \mathcal{W}^*_j)$ incurred by treating unit $j$ as the treated unit and estimating $Y_{jT^*}(0)$ with the placebo SC estimator. Then, fixing some fraction $\alpha \in [0, 1]$, $\mathcal{T}_{T^*}^{B_{\ceil{(1-\alpha) J}}}$ contains the set of plausible treatment effects under the assumption that it is ``no harder'' to estimate $Y_{1T^*}(0)$ when unit $1$ is the treated unit than it is to estimate $Y_{jT^*}(0)$ for any of the $\ceil{(1-\alpha)J}$ ``easiest-to-estimate'' control units, i.e. those with the $\ceil{(1-\alpha)J}$ smallest misspecification error magnitudes. Further, $B_0$ quantifies the magnitude of the misspecification error the SC method would have to incur for a treatment effect of zero to be plausible. This magnitude can be compared to control units' misspecification error magnitudes $B_j$ to benchmark how ``reasonable'' a treatment effect of zero might be, as measured by the percentage $\nu$ of control units for which $B_j \leq B_0$.

Despite the resemblance of our sensitivity analysis to frequentist statistical inference procedures, we caution against interpreting $\nu$ as the $p$-value corresponding to a test of no treatment effect and $\mathcal{T}_{T^*}^{B_{\ceil{(1-\alpha) J}}}$ as a confidence interval for the treatment effect since our methodology is based on the perspective that uncertainty in SC estimates is the result from modeling error, not statistical noise. We believe this perspective is important because in most comparative case studies, we only observe a single outcome sample path over a limited number of time periods for each of a small number of heterogeneous units, only one of which is ever treated \citep{abadie2020using}. As a result, any stochastic model with enough structure to allow for tractable statistical inference in such settings must rely on potentially unrealistic assumptions about the data generating process to make any progress, e.g. distributional assumptions on the stochastic outcome processes, a stance on the treatment assignment mechanism, and/or growing dataset asymptotics.\footnote{\cite{bojinov2019time} and \cite{rambachan2019econometric} discuss similar philosophical issues in the context of time series.}

Further, while some of the statistical approaches to characterizing uncertainty in SC estimates do acknowledge and accommodate the possibility of misspecification error \citep{chernozhukov2017exact,chernozhukov2018practical,cattaneo2019prediction}, the assumptions they make to limit its effect on inferential validity can be difficult to justify in comparative case study settings and interpret for practitioners, e.g. stationarity of units' outcome processes, large numbers of observed pre and post-treatment periods, exchangeability of SC residuals across periods, and/or mean-zero post-treatment SC residuals. While our sensitivity analysis avoids the statistical perspective on estimate uncertainty that is the norm in empirical economics, we believe it provides a transparent evaluation of the credibility of SC counterfactuals in the presence of misspecification error.

\subsubsection{Geometric Motivation for Placebo Tests}

Our methodology also provides an alternative motivation for a variant of the popular design-based placebo test of no treatment effect originally proposed in \cite{abadie2010synthetic}. \cite{abadie2010synthetic} suggest comparing the absolute SC residual $\abs{\hat{R}_{1T^*}^\text{sc}} \coloneqq \absfit{\mathbf{Y}_{0T^*}^T\mathbf{w}_\text{sc} - Y_{1T^*}}$ under the assumption of no treatment effect (so $Y_{1T^*} = Y_{1T^*}(1) = Y_{1T^*}(0)$) to the distribution of absolute placebo residuals $\abs{\hat{R}_{jT^*}^\text{sc}}$ for $j \in \mathcal{J}$; \cite{abadie2010synthetic} interpret $\abs{\hat{R}_{1T^*}^\text{sc}}$ being large relative to $\abs{\hat{R}_{jT^*}^\text{sc}}$ for $j \in \mathcal{J}$ as strong evidence of a non-zero treatment effect, assuming pre-treatment fit is also good. In particular, if we take a design-based perspective and treat outcomes as fixed quantities (see \cite{imbens2015causal}), then under the admittedly unrealistic assumption that treatment is assigned uniformly at random to the units under consideration, the percentage of absolute residuals $\abs{\hat{R}_{jT^*}^\text{sc}}$ that are smaller than $\abs{\hat{R}_{1T^*}^\text{sc}}$ can be interpreted as a $p$-value for a test of the null hypothesis of no treatment effect. \cite{abadie2010synthetic}, \cite{firpo2018synthetic}, and others suggest using test statistics based on the ratios of post-treatment mean squared error under the null hypothesis to pre-treatment prediction error, but in light of the discussion about the relationship between pre and post-treatment error in Section \ref{sec:intro}, it is unclear how meaningful such relative error metrics are in practice.

To see the connection between the placebo test described above and our proposed procedure, recall that the statistic $\nu$ defined at the end of Section \ref{sec:bound_calibration} is computed by asking what fraction of control units' placebo distances $B_j = d_2(\mathbf{w}_\text{sc}^{(j)}, \mathcal{W}_j^*) = \abs{\hat{R}_{jT^*}^\text{sc}}/\normnofit{\mathbf{Y}_{(-j)T^*}}_2$ (from \eqref{eq:def_dist_to_opt}) are smaller than the minimum bound $B_0 = \abs{\hat{R}_{1T^*}^\text{sc}}/\normnofit{\mathbf{Y}_{0T^*}}_2$ (from \eqref{eq:min_dist_expression}) on $d_2(\mathbf{w}_\text{sc}, \mathcal{W}_1^*)$ required for $0$ to lie in the set of plausible treatment effects $\mathcal{T}_{T^*}^{B_0}$. If we multiply $B_0$ and $B_j$ for $j \in \mathcal{J}$ by $\normnofit{\mathbf{Y}_{0T^*}}_2$, we can see that $\nu$ can equivalently be computed by asking for what fraction of control units $j \in \mathcal{J}$ is $B_j\normnofit{\mathbf{Y}_{0T^*}}_2 = \abs{\hat{R}_{jT^*}^\text{sc}} \cdot \normnofit{\mathbf{Y}_{0T^*}}_2/\normnofit{\mathbf{Y}_{(-j)T^*}}_2$ smaller than $B_0\normnofit{\mathbf{Y}_{0T^*}}_2 = \abs{\hat{R}_{1T^*}^\text{sc}}$. Since the ratios $N_j = \normnofit{\mathbf{Y}_{0T^*}}_2/\normnofit{\mathbf{Y}_{(-j)T^*}}_2$ are all greater than one from the discussion at the end of Section \ref{sec:bound_calibration}, we can see that under the assumption of random treatment assignment, $\nu$ can be interpreted as the $p$-value corresponding to a more conservative variant of \cite{abadie2010synthetic}'s placebo test described above. Further, for any $\alpha \in [0, 1]$, we can view $\mathcal{T}_{T^*}^{B_{\ceil{(1-\alpha) J}}}$ as the set of treatment effects under which our conservative version of \cite{abadie2010synthetic}'s placebo test would fail to reject the null hypothesis of zero treatment effect at level $\alpha$. Per the discussion at the end of Section \ref{sec:bound_calibration}, the degree of conservativeness of this placebo test also decreases in the size of the donor pool under mild conditions. Thus, our procedure motivates comparing the treated and control units' absolute residuals to assess errors in SC estimates without starting from a random treatment assignment assumption.

\subsubsection{Donor Pool Selection}

It is important to note that, like most papers in the SC literature, our method assumes the donor pool is fixed before treatment effect estimation. In practice however, researchers often exercise tremendous discretion in donor pool selection in ways that can dramatically change results, as demonstrated by the popular leave-unit-out robustness check we illustrate in Section \ref{subsec:other_robustness}. Despite this sensitivity, inclusion of only the control units that are believed to be ``most similar'' to the treated unit is explicitly advocated for in the SC literature \citep{abadie2020using}. Doing so is encouraged because, as discussed in Footnote \ref{foot:mult_solns}, synthetic controls that interpolate more between control units that are very different from the treated unit can be quite biased if the relationship between pre-treatment outcomes (and covariates) and post-treatment outcomes is non-linear \citep{abadie2020using,crest2018penalized,kellogg2020combining}.

Since our sensitivity analysis defines robustness relative to SC performance when predicting control units' outcomes and the researcher has significant latitude to select those control units, one might worry that our sensitivity analysis is itself sensitive to the choice of donor pool. Because the inclusion or exclusion of a control unit from the donor pool has the potential to affect both the SC estimates of the treated and placebo treated units and the placebo misspecification errors incurred by the SC method, the impact of donor pool manipulation is often ambiguous. It is possible though that an adversarial researcher could select the donor pool to maximize perceived robustness of their SC estimates, but such doctoring has always been a vulnerability of both the SC literature and empirical economics more broadly \citep{meager2020}. 

We note that our procedure \emph{can} assess sensitivity to the inclusion or exclusion of control units that exist in the observed donor pool since such choices are equivalent to toggling the weights corresponding to certain control units between zero and non-zero values. However, we cannot determine the impact of including potential control units not reported by the researcher. For this reason, it is crucial that researchers are transparent about the universe of possible control units from which they select their donor pool and precise about the procedure according to which such selection occurs.

Unfortunately, much ambiguity remains about how researchers should go about defining such a universe. For example, in the context of the tobacco control program studed in \cite{abadie2010synthetic}, one might argue that California is more similar along many dimensions (e.g. total population or GDP) to countries like Germany and the UK than US states like Nebraska or Utah; perhaps data on control units from \cite{abadie2015comparative}'s study of German reunification (augmented with data on tobacco sales) would yield better SC counterfactuals? While such a line of reasoning is compelling, a researcher could also argue that cultural norms around smoking in California are more similar to those of other US states than European countries. While contrived, this small example illustrates the kind of subjectivity inherent in the donor pool selection process, and to our knowledge, there exist no agreed-upon best practices or formal criteria for inclusion or exclusion of particular units. As such, we view studying the effect of donor pool selection on SC estimates with more analytical precision as an important area for future investigation.

\section{Case Studies}\label{sec:case_studies}

\subsection{Applications}

We now demonstrate how to apply our sensitivity analysis as outlined by Procedure \ref{alg:weight_space_sensitivity_analysis} by re-examining three canonical policies studied often in the SC literature: California's tobacco control program on tobacco sales using data provided by \cite{abadie2010synthetic}, German reunification on GDP using data from \cite{abadie2015comparative}, and the Mariel boatlift and Cuban mass migration on the 20th percentile of the wage distribution in Miami using data as in \cite{mariel2019}.

\begin{figure}[p]
	\centering
	\captionsetup{width=\textwidth}
	\begin{subfigure}{\textwidth}
	\centering{}
	\includegraphics[width=0.8\textwidth]{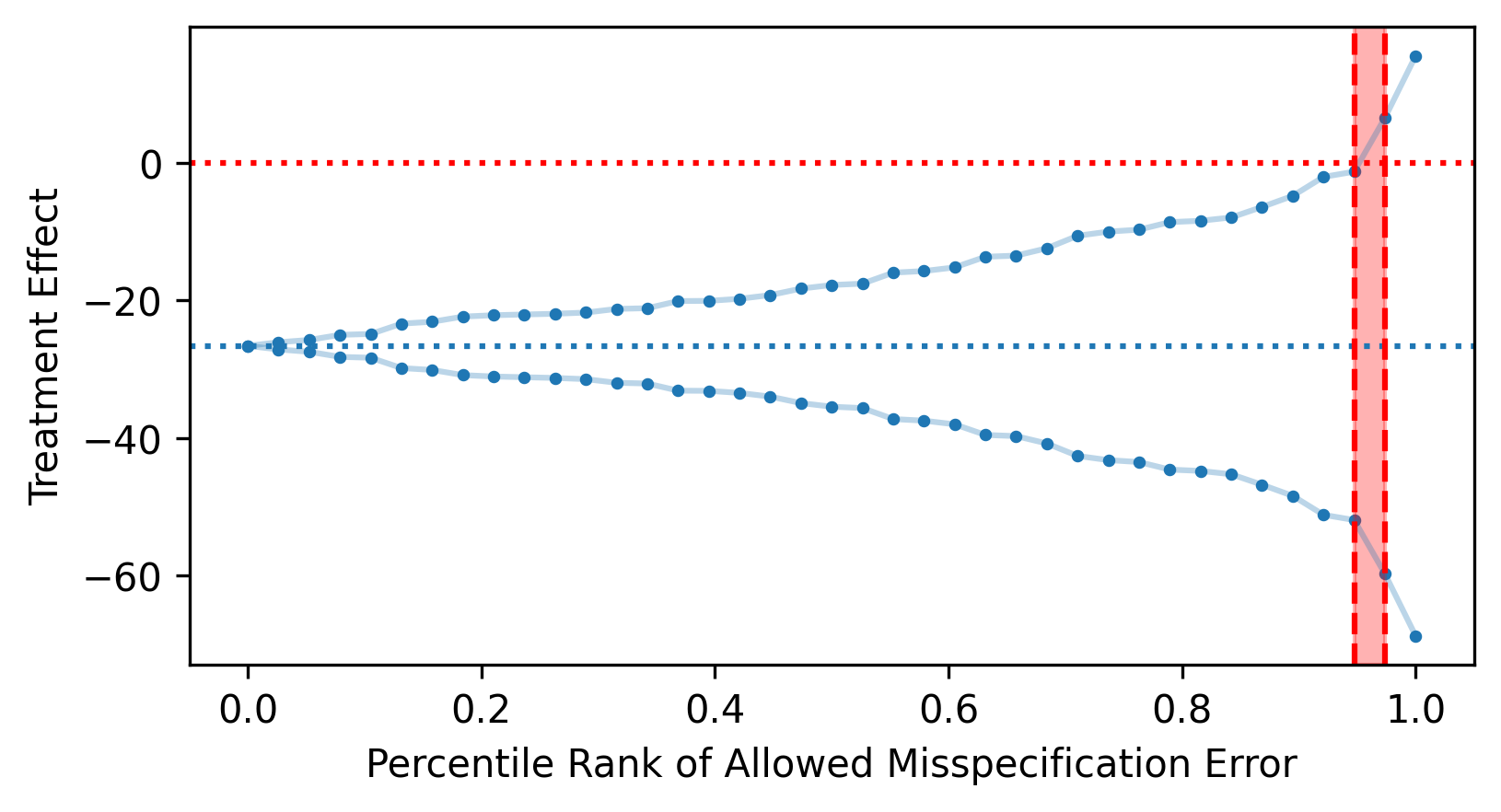}
	\caption{Effect of California's Tobacco Control Program}
	\label{fig:california_quantile_bounds_plot}
	\end{subfigure}
	\begin{subfigure}{\textwidth}
	\centering
	\includegraphics[width=0.8\textwidth]{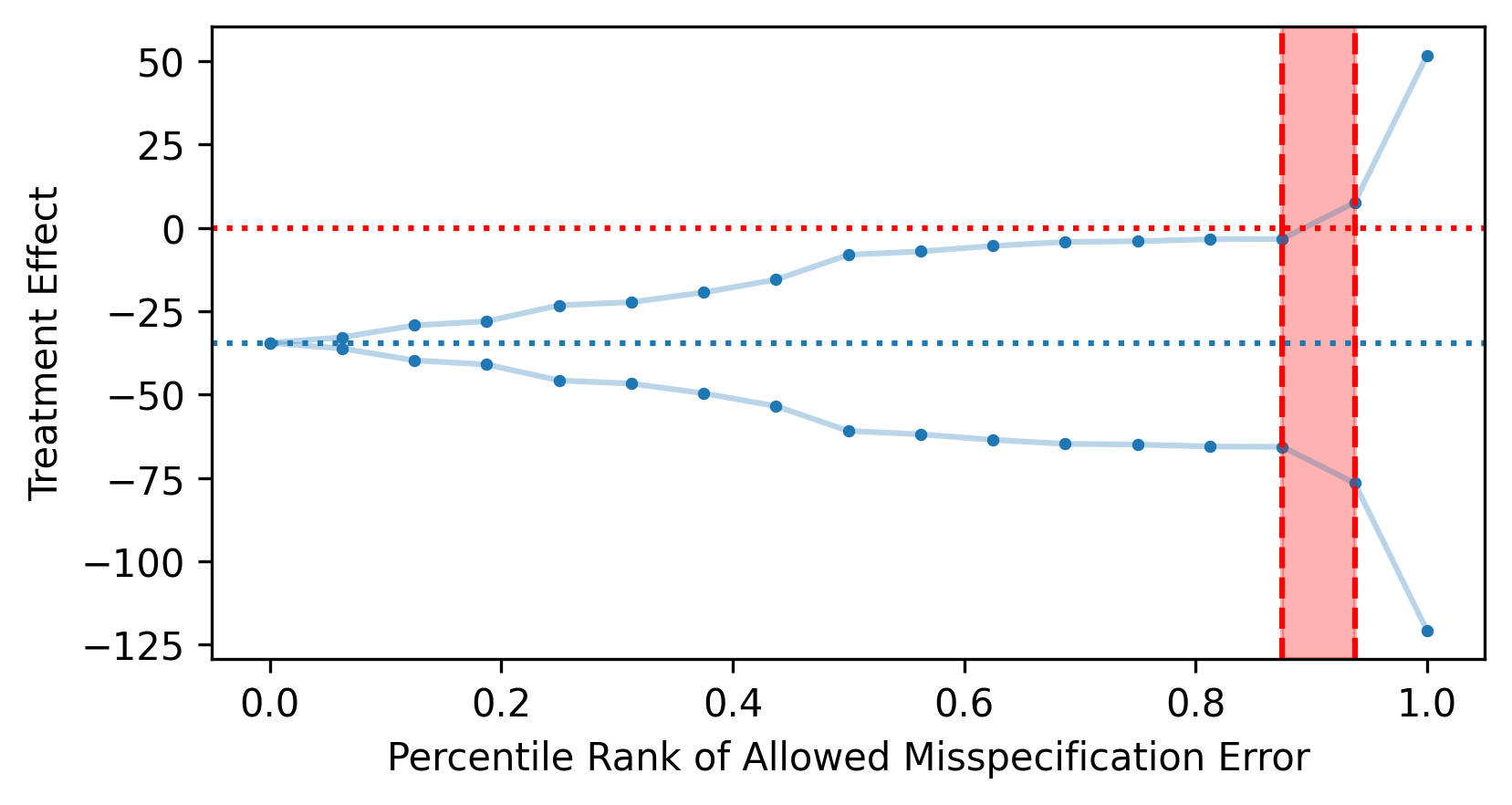}
	\caption{Effect of German Reunification on GDP}
	\label{fig:german_quantile_bounds_plot}
	\end{subfigure}
	\begin{subfigure}{\textwidth}
	\centering
	\includegraphics[width=0.8\textwidth]{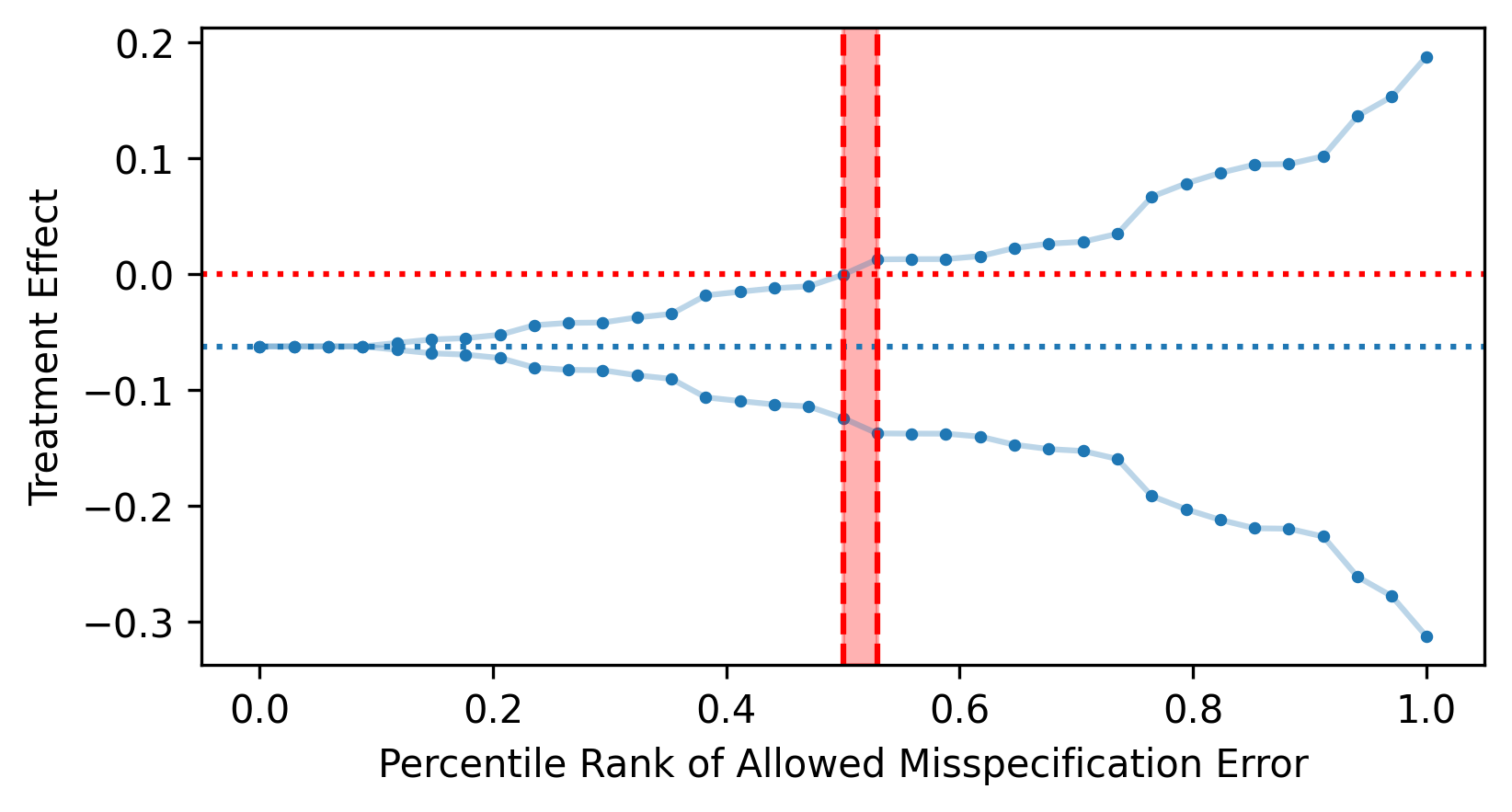}
	\caption{Effect of Mariel Boatlift on Low-Income Wages}
	\label{fig:mariel_boatlift_bounds_plot}
	\end{subfigure}
	\caption{Plots of treatment effect bounds $\mathcal{T}_{T^*}^{B_j}$ corresponding to each control unit $j$'s misspecification error computed using data from three papers using the SC method, where the units of the $x$-axes are percentile ranks $p_j$ of the set of misspecification errors $\{B_j \setst j \in \mathcal{J}\}$. We highlight the region between $B_{j_0}$ and $B_{j_0+1}$ in which our treatment effect bounds first contain zero in red.}
	\label{fig:case_study_plots}
	\end{figure}

In Figure \ref{fig:case_study_plots}, we summarize the results from each case study by plotting the range of possible treatment effects at each percentile rank $p_j = (j-1)/J$ of the ordered set of placebo misspecification errors $\{B_j: j \in \mathcal{J}\}$. In Figure \ref{fig:california_quantile_bounds_plot}, the horizontal, dotted blue line represents the SC point estimate of the effect of California's tobacco program on tobacco sales. For each of the $J$ observed placebo misspecification errors $B_j$, we use blue points to denote the maximum and minimum treatment effects possible for California if we allow for misspecification error up to $B_j$. The $x$-axis represents $B_j$ with its percentile rank $p_j$ within the ordered set of placebo misspecification errors. We highlight in red the interval of the placebo misspecification error distribution where the allowable misspecification error first yields treatment effect bounds containing zero. Summarizing Figure \ref{fig:california_quantile_bounds_plot}, we can see that the SC weight estimates for California would need to incur at least as much error as the 94.7th percentile of the 38 placebo misspecification errors for a zero treatment effect to be plausible. As such, we conclude that this California treatment effect is robust to misspecification error.

In Figure \ref{fig:german_quantile_bounds_plot} we depict the analogous plot for the treatment effect of German reunification on the country's GDP. The effect is slightly less robust to misspecification, as the SC weights for Germany would need to have more misspecification error than 14 (87.5\%) of the placebo treated units. Bounds on the effect of the Mariel boatlift on the 20th percentile of wage distribution in Miami are shown in Figure \ref{fig:mariel_boatlift_bounds_plot}; allowing for the median placebo misspecification error amongst the control units is enough to yield bounds on the treatment effect that contain zero. Since Miami's SC weights would only need to be as incorrect as the median control unit for the sign of the treatment effect to be ambiguous, we bolster \cite{mariel2019}'s conclusion that the small negative treatment effect of the Mariel boatlift on low-income wages purported by \cite{borjas2017wage} is not robust and can be explained by weight misspecification. 

\subsection{Other Robustness Checks}\label{subsec:other_robustness}

Of course, our procedure is not the first to purport to help researchers assess the susceptibility of their SC treatment effect estimates to misspecification error. We next show that our procedure provides more complete and interpretable measures of SC estimate robustness compared to two commonly used robustness checks in the SC literature. In the spirit of \cite{bertrand2004}, we believe methods are best tested on real datasets, so we implement two popular alternative procedures in repeated placebo versions of each of our three case studies, treating each of the control units as a placebo treated unit and comparing the results of these methods to those delivered by our procedure.

\begin{figure}[t]
	\captionsetup{width=0.93\textwidth}
	\begin{subfigure}{0.5\textwidth}
	\centering
	\includegraphics[width=\textwidth]{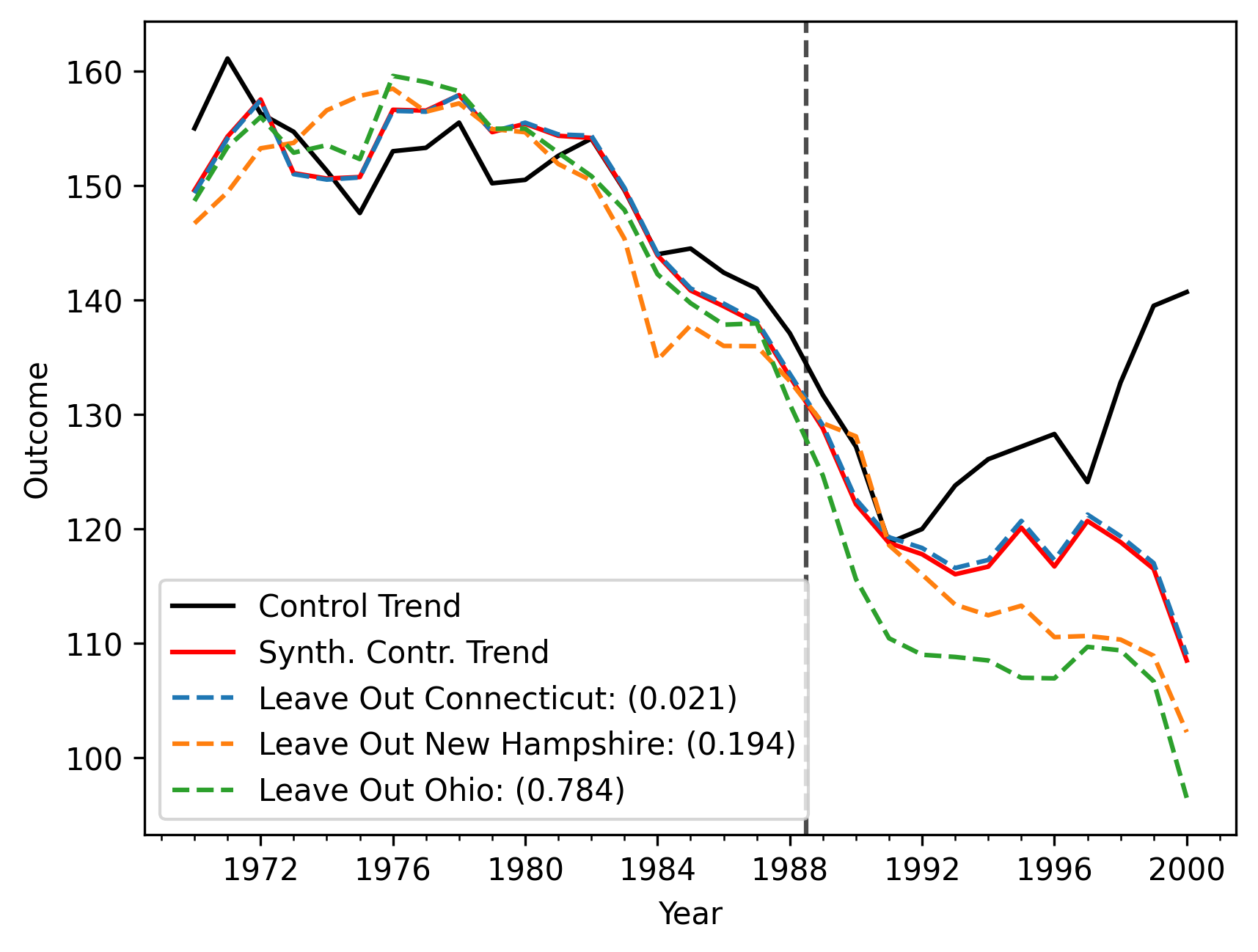}
	\caption{Delaware Leave-Unit-Out Analysis}
	\label{fig:leave_one_out}
	\end{subfigure}
	\begin{subfigure}{0.5\textwidth}
	\centering
	\includegraphics[width=\textwidth]{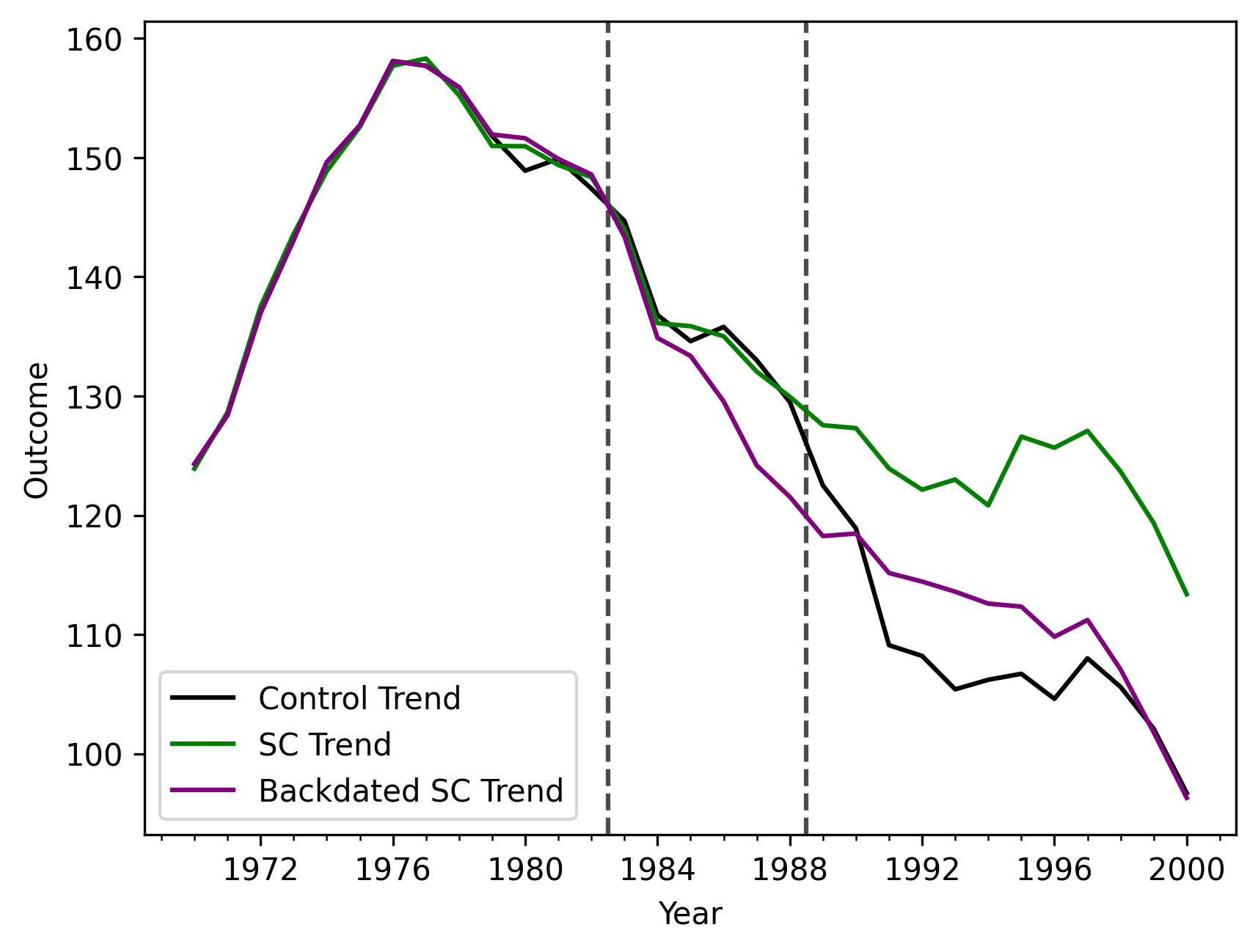}
	\caption{Virginia Leave-Time-Out Analysis}
	\label{fig:leave_time_out}
	\end{subfigure}
	\caption{In Figure \ref{fig:leave_one_out}, we show the SC trends for Delaware computed while leaving out each donor unit that received positive weight (weight shown in parentheses) when running the SC method with the entire donor pool. In Figure \ref{fig:leave_time_out}, we show the SC trend for Virginia computed while excluding the pre-treatment fit errors for the six periods immediately prior to treatment (between the two vertical black lines) from the SC objective.}
	\label{fig:leave_data_out}
	\end{figure}

First, we examine the ``leave-unit-out'' robustness check, which entails dropping each control unit from the donor pool and recomputing SC outcome estimates with a donor pool consisting of the remaining control units \citep{abadie2020using}.\footnote{It suffices to drop only those control units with positive weight in the full-sample vector of SC weights because dropping units with zero weight will not affect SC estimates.} The researcher is then supposed to assess robustness qualitatively by checking whether the set of treatment effects outputted by this procedure have the same signs as and similar magnitudes to the effects computed using the full donor pool. When we treat Delaware as the placebo treated unit in the context of the state-by-state smoking data from \cite{abadie2010synthetic} and conduct the leave-unit-out robustness check, we see that the alternative predictions generated by this procedure fail to capture the extent of the prediction error incurred by the SC method, as illustrated in Figure \ref{fig:leave_one_out}.

Unfortunately, this inadequacy is not isolated to Delaware or the state-level smoking data from \cite{abadie2010synthetic}. If we repeat this placebo procedure with each of the other control units in each of our three case studies, we find that $29$ of the $38$ control units ($76.3\%$) from \cite{abadie2010synthetic}; $11$ of the $16$ control units ($68.7\%$) from \cite{abadie2015comparative}; and $21$ of the $34$ control units ($61.8\%$) from \cite{mariel2019} have last period outcomes outside the range of their corresponding leave-unit-out predictions. In some sense, this result is not so surprising, since the leave-unit-out analysis only assesses the sensitivity of SC estimates to a particular cause of misspecification error: mistakenly including a particular unit in the donor pool and placing positive weight on that unit's outcome in a SC estimate.

The second diagnostic we consider, the ``leave-time-out'' or ``backdating'' procedure, involves fitting a synthetic control using only the pre-treatment outcomes up to some number of periods before the first treatment period; the remaining pre-treatment periods in which control outcomes for the treated unit are known are used as a validation set to assess the quality of the SC method's predictions out-of-sample \citep{abadie2020using}. Treating Virginia as the placebo treated unit in the context of the state-by-state smoking data from \cite{abadie2010synthetic}, we leave out the six time periods before California was treated (between the two black vertical lines in Figure \ref{fig:leave_time_out}) and fit the synthetic control on the remaining pre-treatment periods. Given the gap in Figure \ref{fig:leave_time_out} between the true control trend in black and the backdated synthetic control in purple over the five validation periods, many researchers would be skeptical about their SC estimates. In Virginia's case though, the backdated SC trend predicts the outcome in 2000 remarkably well and clearly outperforms the non-backdated SC trend in the other post-treatment periods.

If a researcher only considers the backdating exercise as a diagnostic for the credibility of the original (non-backdated) SC estimates, then the poor predictive performance of the backdated SC counterfactual in the validation periods correctly indicates that the original SC counterfactual does not reflect the true control trend in the post-treatment periods. However, some researchers also use the backdated SC trend itself to compute treatment effect estimates since the leave-time-out exercise directly tests the predictive performance of the same counterfactual on which treatment effect estimates are based. If Virginia were the treated unit, such researchers would be mislead; the poor fit in the validation periods does not translate into meaningfully subpar post-treatment fit.

\renewcommand{\arraystretch}{1.25}
\begin{table}[t]
	\centering
	\begin{tabular}{clcc}
	\hline 
	Counterfactual &  & California & Germany\\
	\hline 
	\hline 
	 & Donor Pool Size & 38 & 16\\
	\hline 
	\multirow{3}{*}{SC} & False Pos. Rate & 10.5\% & 25.0\%\\
	 & False Neg. Rate & 26.3\% & 12.5\%\\
	 & Total Err. Rate & 36.8\% & 37.5\%\\
	\hline 
	\multirow{3}{*}{Backdated SC} & False Pos. Rate & 13.2\% & 18.7\%\\
	 & False Neg. Rate & 10.5\% & 6.2\%\\
	 & Total Err. Rate & 23.7\% & 24.9\%\\
	 \hline
	\end{tabular}
	\caption{This table summarizes the results from our placebo analyses of the leave-time-out robustness check. A false positive occurs when the backdated SC trend fits well in the validation periods but the counterfactual control trend does not fit well post-treatment; similarly, a false negative occurs when the backdated SC trend does not fit well in the validation periods but the counterfactual control trend does fit well post-treatment. The error rate is defined as the sum of the false positive and false negative rates. We report error rates for both backdated and non-backdated SC counterfactual trends.}
	\label{table:leave_time_out_placebo_results}
\end{table}

Given these concerns, we repeat this placebo analysis with each of the other control units in the studies of California's tobacco control law and German reunification.\footnote{Unfortunately, there are not enough pre-treatment periods in the data from \cite{mariel2019} to reliably evaluate SC predictions from the backdated fit.} In particular, we visually code each application of the backdating procedure as yielding a ``false positive''---the backdated SC trend fits well in the validation periods but the counterfactual control trend does not fit well post-treatment---a ``false negative''---the backdated SC trend does not fit well in the validation periods but the counterfactual control trend does fit well post-treatment---or neither if the procedure properly rejected a counterfactual with bad post-treatment fit or did not reject a counterfactual with good post-treatment fit.\footnote{Our notions of good and poor fit here are necessarily heuristic, since we know of no accepted formal criteria in the literature for what constitutes acceptable fit in the validation periods. A more systematic way to code each placebo analysis would be to survey a sample of practitioners who use the SC method and ask them whether they find the predictive performance of backdated and non-backdated SC trends acceptable; we leave such a survey for future work.} Since there is not a consensus amongst practitioners about which of the backdated or non-backdated SC trends should determine treatment effect estimates, we report false positive and false negative rates for both types of counterfactuals.

As can be seen from Table \ref{table:leave_time_out_placebo_results}, while the performance of the leave-time-out procedure is better than the performance of the leave-unit-out procedure, it still leaves much to be desired given that it had the potential to mislead researchers roughly a quarter of the time it was applied in our placebo analyses. Again, these results should not be unexpected, since the leave-time-out analysis is only assessing the sensitivity of SC estimates to misspecification error caused by overfitting to outcomes close to the first treated period. More importantly, there are no agreed-upon formal criteria we know of in the literature for trusting or doubting synthetic control estimates based on the leave-unit-out or backdating exercises. Researchers (including us) seem to decide based on visual appeal, which, as we have demonstrated, can lead researchers astray. In fact, we could not reproduce the error rates in Table \ref{table:leave_time_out_placebo_results} when we conducted the coding exercise described above twice, six months apart; the version included here reports the results from our second coding attempt, and departures from our first results were not uniform in any direction. In contrast, our procedure provides a comprehensive and less subjective approach to assessing how all types of misspecification error could affect SC estimates.

\begin{figure}[t!]
	\centering
	\captionsetup{width=\textwidth}
	\includegraphics[width=0.65\textwidth]{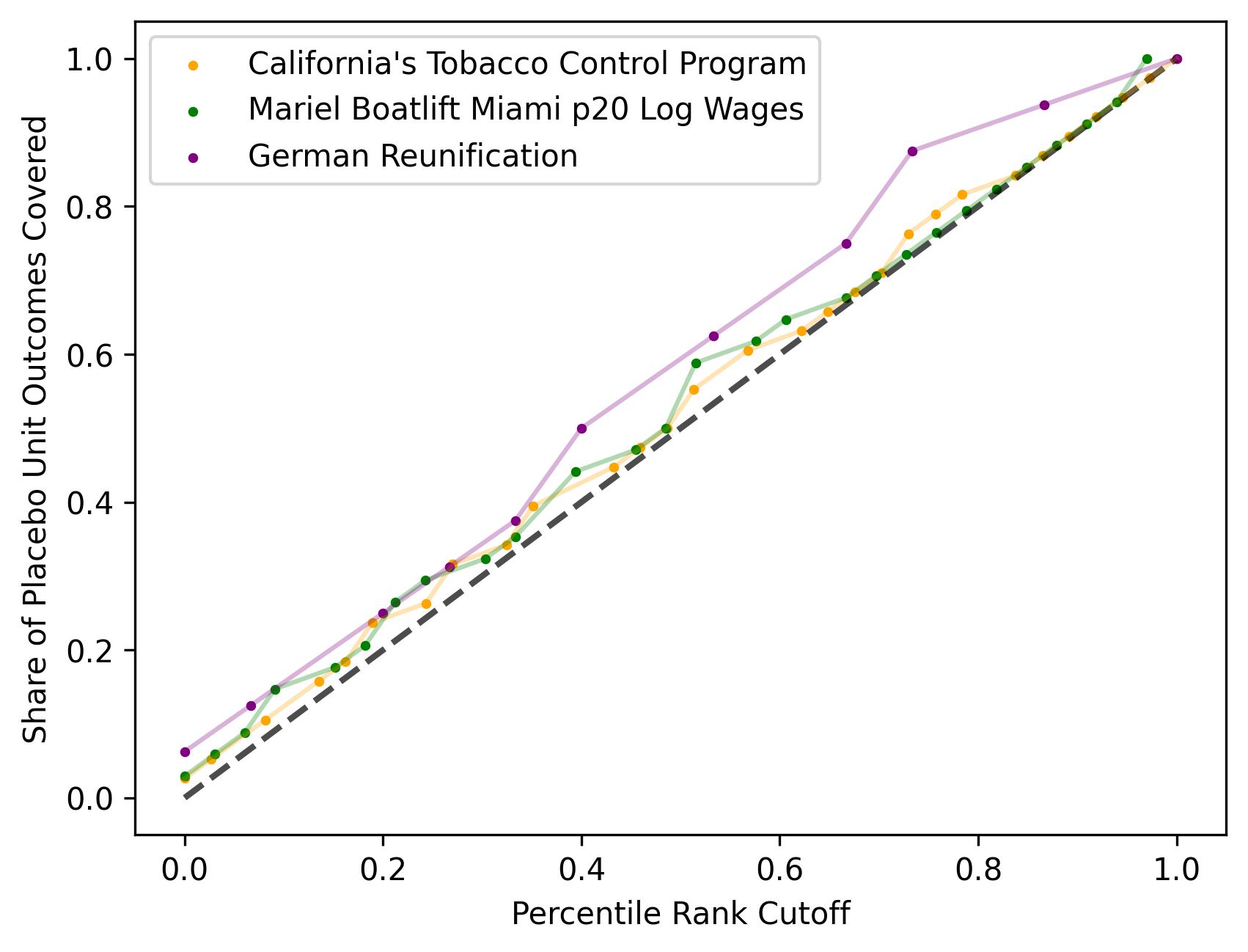}
	\caption{We plot the results from placebo analyses of Procedure \ref{alg:weight_space_sensitivity_analysis} using data from three case studies. On the $x$-axis, we vary the percentile rank cutoff that determines the width of the bounds generated by our procedure and on the $y$-axis we show the share of placebo treated units for which those bounds correctly include a zero treatment effect. The dashed black 45-degree line demarcates $p$ percent of placebo units' true outcomes being covered with a misspecification error cutoff at the $p$th percentile rank.}
	\label{fig:case_study_weight_space_placebo}
	\end{figure}

To assess the effectiveness of our proposed method in comparison, we subject it to the same placebo analyses we used to interrogate the leave-unit-out and leave-time-out robustness checks studied above. In particular, we treat the control units in each of our three case studies as placebo treated units and apply our sensitivity analysis to each. In Figure \ref{fig:case_study_weight_space_placebo}, we plot the share of control units for which our procedure yields bounds on the treatment effect that correctly contain zero for each possible percentile rank at which we could generate bounds using our procedure. When the researcher chooses a threshold of misspecification error in terms of a $p$th percentile rank cutoff that they deem ``acceptable'' when constructing treatment effect bounds, our placebo analyses suggest that doing so correctly captures zero treatment effects for approximately $p$ percent of the placebo treated units. In some sense, this calibrated relationship between the chosen percentile rank cutoff and bound coverage of placebo treated units' outcomes is not surprising, since our procedure can be viewed as a particular way of synthesizing the results of repeated placebo analyses. Section \ref{sec:placebo_analysis} of the Appendix describes the mechanics behind this connection in more detail.

In contrast, it is difficult to translate the results of the leave-unit-out and leave-time-out robustness checks into clear insights about the validity of the SC treatment effect estimate for the treated unit. Recall the leave-unit-out placebo analyses conducted using the data from \cite{abadie2010synthetic}; if the true control outcomes lie outside of the range of leave-unit-out trends for $76.3\%$ of placebo treated units, are we meant to believe that the range of leave-unit-out trends for California only captures the true counterfactual control trend $23.7\%$ of the time under some sampling model for the potential outcomes? As discussed above, it is even harder to understand how informative the backdating placebo analyses are about the value of the leave-time-out analysis applied to the treated unit. While these other methods are plagued by ambiguities in implementation and interpretation, the only degree of freedom left by our procedure for the researcher to determine, the acceptable percentile rank cutoff, is both directly meaningful and closely connected to a natural summary statistic of our procedure's performance in placebo analyses.

\section{Other Misspecification Error Metrics}\label{subsec:misspec_generalize}

\subsection{A Generalized Sensitivity Analysis}\label{subsec:gen_sensitivity_analysis}

The $\ell_2$-distances defined in Section \ref{sec:sensitivty_analysis} between the SC weights and the closest weights that correctly predict $Y_{jT^*}(0)$, $d_2\left(\mathbf{w}_\text{sc}, \mathcal{W}^*_1\right)$ and $d_2(\mathbf{w}_\text{sc}^{(j)}, \mathcal{W}^*_j)$, are natural measures of misspecification error magnitudes, but they are certainly not the only ones researchers can use to assess the sensitivity of SC treatment effect estimates. Instead of measuring the misspecification error incurred by the SC method relative to weights $\mathbf{w}$ with the $\ell_2$-distance of $\mathbf{w}_\text{sc}^{(j)}$ to $\mathbf{w}$, we can use any function $m_j \colon \R^{J - \ind{j \neq 1}} \rightarrow [0, \infty]$ such that $m_j(\mathbf{w}_\text{sc}^{(j)}) = 0$ to measure the distance of $\mathbf{w}$ to $\mathbf{w}_\text{sc}^{(j)}$. To allow for these alternative misspecification error metrics $m_j$, we generalize our proposed sensitivity analysis in Procedure \ref{alg:general_sensitivity_analysis}, which nests the analysis described in Section \ref{sec:sensitivty_analysis} for $m_j(\mathbf{w}) = m^\text{wt}_j(\mathbf{w}) \coloneqq \normnofit{\mathbf{w}_\text{sc}^{(j)} - \mathbf{w}}_2$. Note that as long as $m_j$ are convex functions, then although the optimization problems \eqref{eq:min_dist_to_opt_weights_control_unit_general}, \eqref{eq:bounds_on_cntrfct_control_outcomes_general}, and \eqref{eq:nullifying_distance_comp_def_general} likely do not have closed-form solutions as their equivalents in Section \ref{sec:sensitivty_analysis} do, their solutions are still easily computable numerically using off-the-shelf convex optimization software \citep{boyd2004convex}.

\begin{algbox}[p]
\myalg{alg:general_sensitivity_analysis}{Generalized Sensitivity Analysis}{
\begin{enumerate}[leftmargin=3ex, itemsep=0ex]
	\item For each control unit $j \in \mathcal{J}$:
	\begin{enumerate}[itemsep=-0.5ex, parsep=-0.5ex]
		\item Compute the misspecification error $d_{m_j}(\mathbf{w}_\text{sc}^{(j)}, \mathcal{W}_j^*)$ incurred by estimating the placebo post-treatment outcome of interest $Y_{jT^*}(0)$ for control unit $j$ with the SC method using the other $J-1$ units in the donor pool as control units, as in \eqref{eq:min_dist_to_opt_weights_control_unit}:
		\begin{equation}\label{eq:min_dist_to_opt_weights_control_unit_general}
		\begin{aligned}
			d_{m_j}(\mathbf{w}_\text{sc}^{(j)}, \mathcal{W}^*_j) \coloneqq \inf_{\mathbf{w} \in \R^{J-1}}&~ m_j(\mathbf{w}) \\
		\suchthat&~ \mathbf{Y}_{(-j)T^*}^T\mathbf{w} = Y_{jT^*} ~~\left(\Leftrightarrow \mathbf{w} \in \mathcal{W}_j^*\right).
		\end{aligned}
		\end{equation}
		\item Compute the largest and smallest plausible counterfactual control outcomes $Y_{1T^*}^{B_j,-}(0)$ and $Y_{1T^*}^{B_j,+}(0)$ under the assumption that the misspecification error $d_{m_1}(\mathbf{w}_{sc}, \mathcal{W}^*_1)$ incurred by estimating $Y_{1T^*}(0)$ with the SC method is at most the misspecification error $B_j \coloneqq d_{m_j}(\mathbf{w}_\text{sc}^{(j)}, \mathcal{W}^*_j)$, as in \eqref{eq:bounds_on_cntrfct_control_outcomes}:
		\begin{equation}\label{eq:bounds_on_cntrfct_control_outcomes_general}
		\begin{aligned}
			Y_{1T^*}^{B_j,-}(0) &\coloneqq \inf_{\mathbf{w} \in \R^J} \left\{\mathbf{Y}_{0T^*}^T\mathbf{w} 
			~\setst~ m_1(\mathbf{w}) \leq d_{m_j}(\mathbf{w}_\text{sc}^{(j)}, \mathcal{W}^*_j)\right\} \\
			Y_{1T^*}^{B_j,+}(0) &\coloneqq \sup_{\mathbf{w} \in \R^J} \left\{\mathbf{Y}_{0T^*}^T\mathbf{w}
			~\setst~ m_1(\mathbf{w}) \leq d_{m_j}(\mathbf{w}_\text{sc}^{(j)}, \mathcal{W}^*_j)\right\}
		\end{aligned}
		\end{equation}
		\item Compute the bounds $\mathcal{T}^{B_j}_{T^*}$ on the treatment effect $\tau_{T^*}$ under the assumption that the misspecification error $d_{m_1}(\mathbf{w}_{sc}, \mathcal{W}^*_1)$ incurred by estimating $Y_{1T^*}(0)$ with the SC method is at most the misspecification error $B_j$, as in \eqref{eq:def_T_B_j_bounds}:
		\begin{equation}\label{eq:def_T_B_j_bounds_gen}
		\begin{aligned}
			\mathcal{T}^{B_j}_{T^*} &\coloneqq \left[Y_{1T^*} - Y_{1T^*}^{B_j,+}(0), Y_{1T^*} - Y_{1T^*}^{B_j, -}(0)\right]
		\end{aligned}
		\end{equation}
	\end{enumerate}

	\item Compute the minimum misspecification error $B_0$ needed for $0 \in \mathcal{T}^{B_0}_{T^*}$, i.e. $0$ to be a plausible treatment effect estimate, as in \eqref{eq:nullifying_distance_comp_def}:
	\begin{equation}\label{eq:nullifying_distance_comp_def_general}
	\begin{aligned}
	B_0 \coloneqq \inf_{\mathbf{w} \in \R^J}&~ m_1(\mathbf{w}) \\
	\suchthat &~ \mathbf{Y}_{0T^*}^T\mathbf{w} = Y_{1T^*}
	\end{aligned}
	\end{equation}
	and find the control unit $j_0$ with the largest misspecification error still smaller than $B_0$, i.e. where $B_{j_0} \leq B_0 \leq B_{j_0+1}$.
	
	\item Visualize the treatment effect bounds $\mathcal{T}^{B_j}_{T^*}$ for each $j \in \mathcal{J}$ and the misspecification errors $B_{j_0}$ and $B_{j_0 + 1}$ in a plot like Figure \ref{fig:california_quantile_bounds_plot}, and report the percentage $\nu = (j_0 - 1)/J$ of control units whose misspecification errors $B_j$ are smaller than $B_0$.
\end{enumerate}
}
\end{algbox}

To demonstrate the value of this more general procedure, we focus on an alternative misspecification error metric $m^\text{err}_j(\mathbf{w})$, defined as the extra pre-treatment prediction error incurred by $\mathbf{w}$ relative to the minimum achievable pre-treatment prediction error with valid SC weights, assuming $\mathbf{w}$ are also valid SC weights:
\begin{equation}\label{eq:def_m_j_err}
m^\text{err}_j(\mathbf{w}) \coloneqq \frac{\norm{\mathbf{x}_j - X_{-j}\mathbf{w}}_2 + \psi_{\Delta_{J - \ind{j \neq 1}}}(\mathbf{w})}{\min_{\tilde{\mathbf{w}} \in \R^{J - \ind{j \neq 1}}} \left\{\norm{\mathbf{x}_j - X_{-j}\tilde{\mathbf{w}}}_2 + \psi_{\Delta_{J - \ind{j \neq 1}}}(\tilde{\mathbf{w}}) \right\}} - 1,
\end{equation}
where for a given set $C \subseteq \R^{J - \ind{j \neq 1}}$, $\psi_{C}(\mathbf{w})$ is a penalty term designed to constrain $\mathbf{w}$ to lie in the set $C$ when $m_j$ is used in minimization problems:
\begin{equation}\label{eq:constraint_penality}
	\psi_{C}(\mathbf{w}) \coloneqq \begin{cases}
		0 & \mathbf{w} \in C \\
		\infty & \text{otherwise}
	\end{cases}
\end{equation}

The denominator of the fraction in \eqref{eq:def_m_j_err} is just the pre-treatment prediction error incurred by the canonical SC estimator, since $\norm{\mathbf{x}_j - X_{-j}\tilde{\mathbf{w}}}_2$ is exactly the objective function minimized in \eqref{eq:basic_synth_contr_method_placebo} to construct a synthetic control and $\psi_{\Delta_{J - \ind{j \neq 1}}}(\tilde{\mathbf{w}})$ just ensures that the minimizer of $\norm{\mathbf{x}_j - X_{-j}\tilde{\mathbf{w}}}_2 + \psi_{\Delta_{J - \ind{j \neq 1}}}(\tilde{\mathbf{w}})$ is a vector of valid SC weights. Then, if we use $m^\text{err}_j$ as the misspecification error metric in our proposed sensitivity analysis, we can interpret the misspecification error $d_{m^\text{err}_j}(\mathbf{w}_\text{sc}^{(j)}, \mathcal{W}^*_j)$ as the minimum amount of additional pre-treatment prediction error (relative to the minimum possible) a researcher would have to tolerate for a vector of SC weights that yields a correct prediction of $Y_{jT^*}(0)$ to be considered a ``reasonable'' choice of weights.

As it happens, the weights that solve \eqref{eq:min_dist_to_opt_weights_control_unit_general} under $m^\text{err}_j$ are exactly the weights that yield the green outcome trends in Figure \ref{fig:intro_placebo} that match Virginia and Delaware's outcomes in 2000 and achieve the smallest possible pre-treatment prediction error magnitudes while doing so. Further, suppose we treat unit $j$ as the treated unit and the other $J-1$ control units as the donor pool. Then the sets $[Y_{jT^*}^{B_j,-}(0), Y_{jT^*}^{B_j,+}(0)]$ for $j \in \mathcal{J}$ with endpoints defined analogously to \eqref{eq:bounds_on_cntrfct_control_outcomes_general}, i.e. the sets that contain the plausible predicted control outcomes for each unit $j$ assuming misspecification error is no larger than $j$'s own true misspecification error, are exactly the red dashed intervals in Figures \ref{fig:intro_virginia} and \ref{fig:intro_delaware}.

Although $m^\text{err}_j$ has clear intuitive appeal, it does have several shortcomings. First, it is only well-defined if $X_{-j} \mathbf{w} \neq \mathbf{x}_j$ for all $\mathbf{w} \in \Delta_{J - \ind{j \neq 1}}$; otherwise, the denominator in \eqref{eq:def_m_j_err} will be zero, in which case $m^\text{err}_j$ is unusable given the dataset of interest. Second, the sets of $\mathbf{w}$ that perfectly predict the period-$T^*$ outcomes for the control units with the largest and smallest values of $Y_{jT^*}$ do not intersect with $\Delta_{J - \ind{j \neq 1}}$ at all, in which case $m^\text{err}_j(\mathbf{w})$ will be infinite for all feasible $\mathbf{w}$ in \eqref{eq:min_dist_to_opt_weights_control_unit_general}. Then $d_{m_j}(\mathbf{w}_\text{sc}^{(j)}, \mathcal{W}^*_j) = \infty$ for the two units with the largest and smallest period-$T^*$ outcomes, meaning $\mathcal{T}^{B_j}_{T^*} = (-\infty, \infty)$. Despite the fact that these bounds contain the whole real line, we do not intend their vacuousness to reflect that all treatment effects are equally plausible; we simply mean to convey that the particular bounds corresponding to the control units with extreme outcomes are uninformative about the treatment effect for the treated unit.

In addition to the generalization of our sensitivity analysis to other misspecification error metrics described above, we also extend our procedure to measure the sensitivity of alternative outcome contrast estimates in Section \ref{subsec:other_contrasts} of the Appendix and to apply to effect estimates generated by other policy evaluation methods for panel data in Section \ref{subsec:lpes} of the Appendix. Further, we demonstrate in Section \ref{sec:mult_solns_appendix} of the Appendix how this generalized procedure can be used to account for potential non-uniqueness of the SC estimator in the sensitivity analysis from Section \ref{sec:sensitivty_analysis}.

\subsection{Choosing a Misspecification Error Metric}\label{subsec:case_studies_revisited}

To understand how the choice of misspecification error metric can affect the output of Procedure \ref{alg:general_sensitivity_analysis}, we compare the results of our sensitivity analysis based on $m^\text{wt}_j$ shown in Figure \ref{fig:case_study_plots} to results based on two additional misspecification error metrics, which we review below:
\begin{enumerate}[itemsep=0ex]
	\item \emph{Unconstrained weight space}: $m^\text{wt}_j(\mathbf{w}) = \normnofit{\mathbf{w}_\text{sc}^{(j)} - \mathbf{w}}_2$; as described above, using this metric yields the sensitivity analysis given in Procedure \ref{alg:weight_space_sensitivity_analysis}.
	
	\item \emph{Constrained weight space}: $m^\text{wt}_{\Delta_{J - \ind{j \neq 1}}}(\mathbf{w}) \coloneqq \normnofit{\mathbf{w}_\text{sc}^{(j)} - \mathbf{w}}_2 + \psi_{\Delta_{J - \ind{j \neq 1}}}(\mathbf{w})$; this metric still measures distance in weight space but requires $\mathbf{w}$ to lie in the set of valid SC weights.

	\item \emph{Constrained error space}:
	\begin{equation*}
	m^\text{err}_j(\mathbf{w}) \coloneqq \frac{\norm{\mathbf{x}_j - X_{-j}\mathbf{w}}_2 + \psi_{\Delta_{J - \ind{j \neq 1}}}(\mathbf{w})}{\min_{\tilde{\mathbf{w}} \in \R^{J - \ind{j \neq 1}}} \left\{\norm{\mathbf{x}_j - X_{-j}\tilde{\mathbf{w}}}_2 + \psi_{\Delta_{J - \ind{j \neq 1}}}(\tilde{\mathbf{w}}) \right\}} - 1;
	\end{equation*}
	as discussed in Section \ref{subsec:gen_sensitivity_analysis}, this metric measures distance with the extra error incurred by $\mathbf{w}$ relative to the error incurred by the vector of SC weights.
\end{enumerate}

\begin{figure}[p]
	\centering
	\captionsetup{width=\textwidth}
	\begin{subfigure}{\textwidth}
	\centering
	\includegraphics[width=0.8\textwidth]{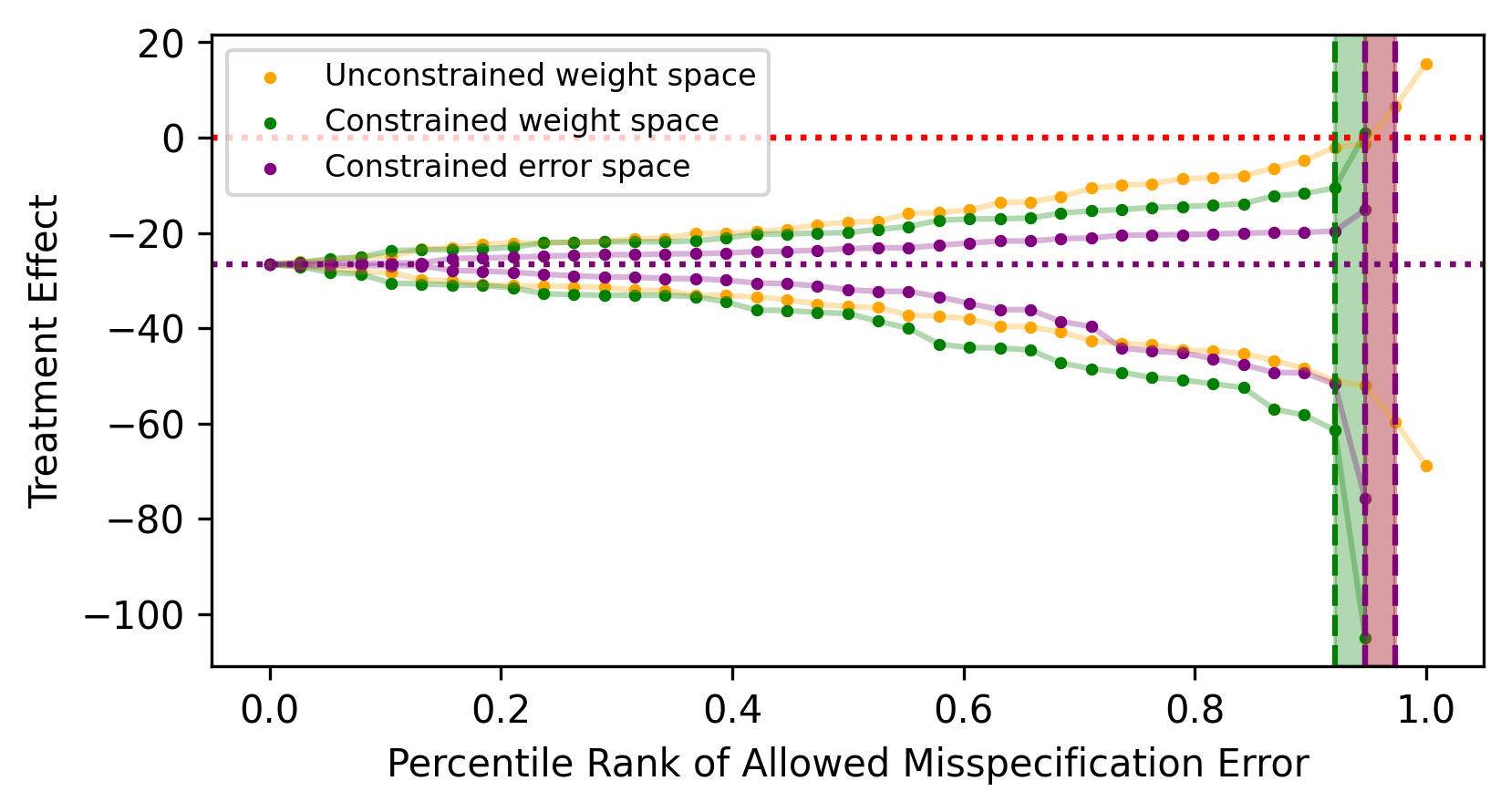}
	\caption{Effect of California's Tobacco Control Program}
	\label{fig:california_compare}
	\end{subfigure}
	\begin{subfigure}{\textwidth}
	\centering
	\includegraphics[width=0.8\textwidth]{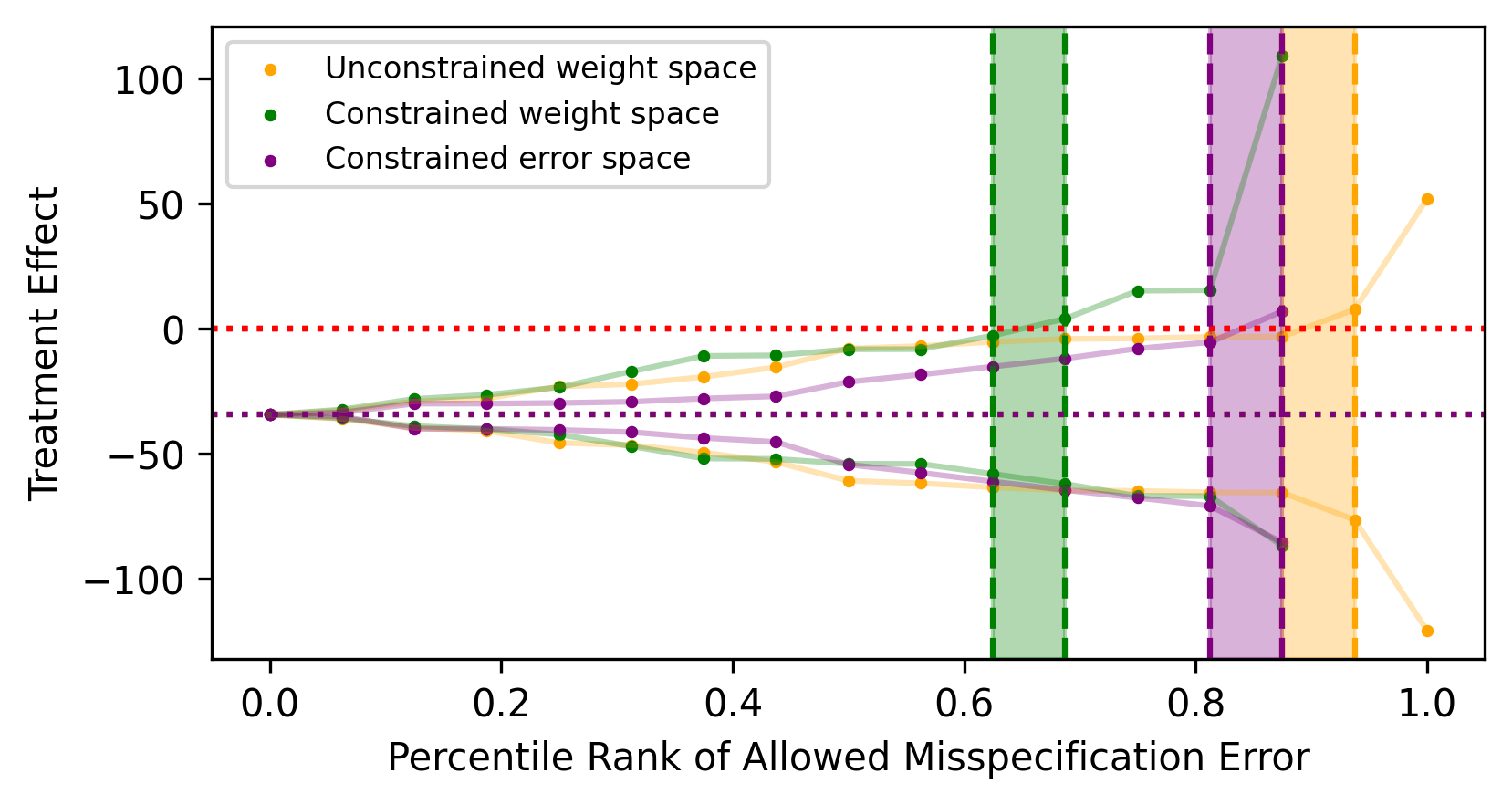}
	\caption{Effect of German Reunification on GDP}
	\label{fig:german_compare}
	\end{subfigure}
	\begin{subfigure}{\textwidth}
	\centering
	\includegraphics[width=0.8\textwidth]{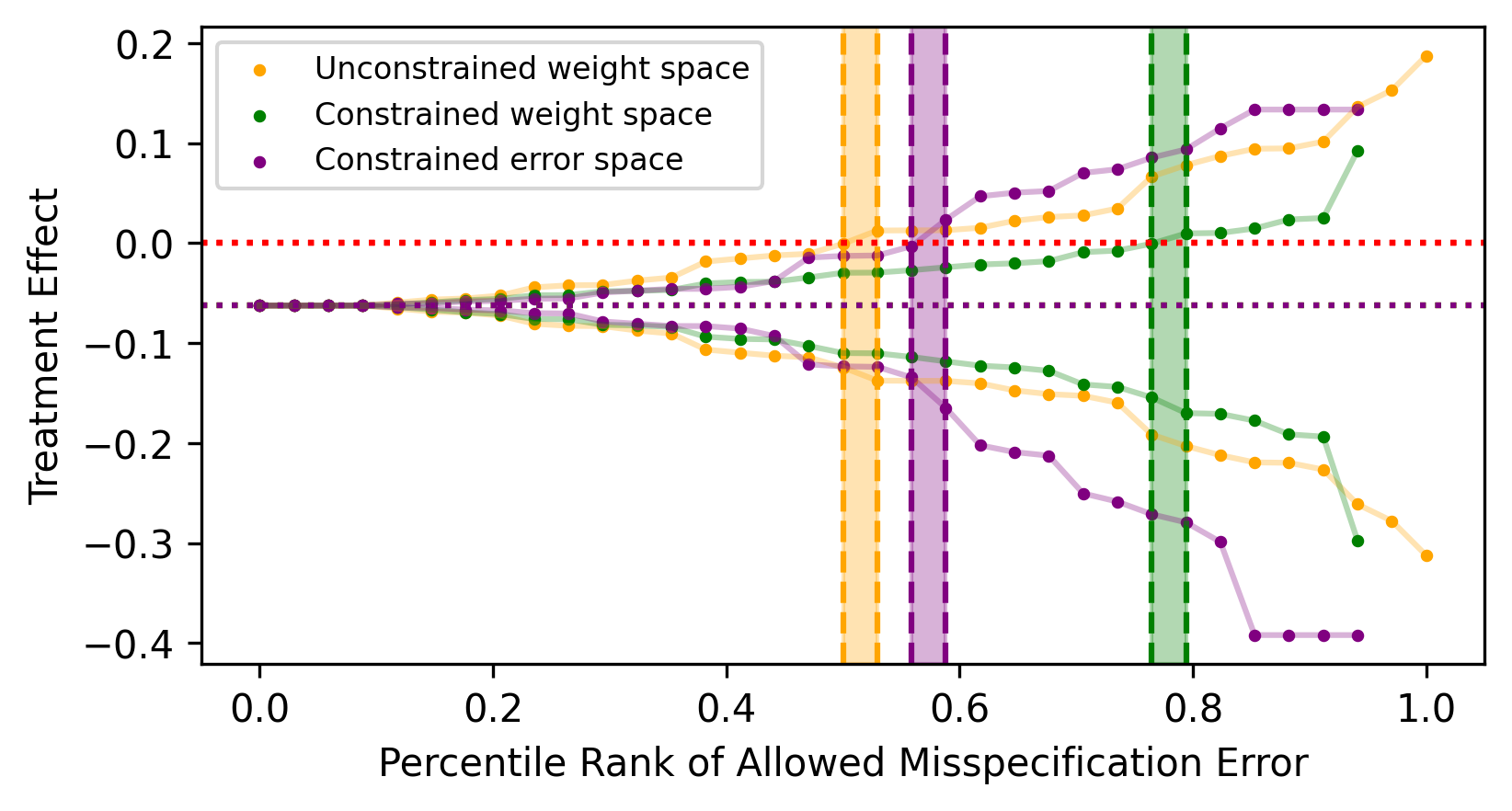}
	\caption{Effect of Mariel Boatlift on Low-Income Wages}
	\label{fig:mariel_boatlift_compare}
	\end{subfigure}
	\caption{Plots of treatment effect bounds $\mathcal{T}_{T^*}^{B_j}$ using different misspecification error metrics corresponding to each control unit $j$'s misspecification error computed using data from three papers using the SC method, where the units of the $x$-axes are percentile ranks $p_j$ of the set of misspecification errors $\{B_j \setst j \in \mathcal{J}\}$. We shade each region between $B_{j_0}$ and $B_{j_0+1}$ in which our treatment effect bounds first contain zero in the color corresponding to the relevant misspecification error metric.}
	\label{fig:case_study_compare}
	\end{figure}

The results of repeating our earlier case studies with the alternative misspecification metrics described above are shown in Figure \ref{fig:case_study_compare}. First, the treatment effect bounds for the tobacco control program in California in Figure \ref{fig:california_compare} uniformly indicate that the finding of a large, negative effect is robust, since for all choices of $m_j$, the misspecification error for California would need to be large relative to the misspecification errors of most control units.  For a zero treatment effect to be plausible, the constrainted weight space metric suggests California's misspecification error would need to be larger than 92.1\% of the control units and the remaining two metrics require error larger than 94.7\% of control units. The results for the German reunification and Mariel boatlift settings shown in Figures \ref{fig:german_compare} and \ref{fig:mariel_boatlift_compare} have more variation across misspecification metrics. While the constrained error space metric suggests fairly robust results in the German reunification setting, the other two are less supportive. In the Mariel boatlift setting, the treatment effect of the influx of immigrants on low-income wages in Miami is reasonably indistinguishable from zero using the unconstrained weight space and constrained error space metrics, as has been argued in the literature by other means.

Perhaps counterintuitively, Figure \ref{fig:german_compare} demonstrates that the percentile rank of the misspecification error needed for a zero treatment effect to be plausible using the constrained weight space metric is smaller than the equivalent percentile rank using the unconstrained weight space metric. At first, this phenomenon may seem impossible since, holding the magnitude of misspecification error fixed, the bounds constructed by maximizing and minimizing over the unconstrained set of weights in \eqref{eq:bounds_on_cntrfct_control_outcomes_general} should be mechanically wider than the bounds constructed over the constrained set. However, recall that the units on the $x$-axis in \ref{fig:german_compare} correspond to the \emph{percentile ranks} of the placebo misspecification errors, not their magnitudes. Because the relative sizes of the placebo misspecification errors also depend on choice of metric, it is certainly possible that, at a fixed percentile rank in the placebo misspecification error distribution, either metric could yield wider bounds. While this ambiguity may suggest visualizing the bounds defined via the two weight space metrics in terms of absolute misspecfication error magnitudes, as discussed in Section \ref{sec:bound_calibration}, it is hard to determine what constitutes a reasonable amount of misspecification error measured using $\ell_2$-distances in weight space. Benchmarking against the placebo misspecification errors of the control units provides a more meaningful characterization of the robustness of SC estimates.

Unfortunately, given the ambiguous relationships between metrics discussed above, we cannot recommend a single preferred misspecification error metric for all settings. Rather, we believe the choice should be made based on the researcher's prior beliefs about the SC method's susceptibility to misspecification error. When comparing the constrained and unconstrained weight space metrics, the decision should be determined by the researcher's belief about the validity of the SC weight constraints. If the researcher just views the constraints as a convenient way of inducing sparsity in the SC weights, then conducting the sensitivity analysis while enforcing those constraints would fail to capture the possible misspecification error induced by the imposition of the constraints when choosing the SC weights. However, if the researcher believes the weight constraints capture important structural features of the setting, for example that treatment effect estimates based on extrapolation are undesirable, then they may wish to use the constrained weight space metric and only evaluate misspecification error incurred by the minimization of the wrong objective function when selecting the SC weights, not the weight constraints themselves.\footnote{by extrapolation, we mean estimates of $Y_{1T^*}$ that lie outside the range of control units' period-$T^*$ outcomes \citep{abadie2020using}.}

The choice between the constrained weight and constrained error space metrics is more subtle. The sensitivity analyses based on weight space metrics search for alternative weights agnostic to direction when constructing treatment effect bounds. On the other hand, the constrained error space metric penalizes alternative weights that have poor performance on the original SC objective. Therefore, if the researcher does not believe pre-treatment fit is at all informative about post-treatment fit, they may prefer the weight space metrics. However, if the researcher maintains that good pretreatment fit is a desirable and informative property of the weights used to construct counterfactual predictions, the constrained error space metric may make more sense. In principle, one could even interpolate between the different metrics.

\section{Conclusion}\label{sec:discussion}

In this paper, we  demonstrate that pre-treatment fit is neither neccesary nor sufficient for good post-treatment fit and that existing robustness checks often fail to capture the extent of this disconnect due to their heuristic motivations and ad-hoc interpretations. To structure conversations about the robustness of SC estimates, we provide researchers with a procedure to systematically assess SC estimate sensitivity to misspecification error in an interpretable, data-driven manner. Our method can flexibly encode researchers' varying beliefs about the validity of the assumptions made when interpretating SC estimates as causal by accommodating different measures of misspecification error. 

Since it is difficult to determine which statistical models are appropriate for comparative case study settings with small numbers of heterogeneous units observed over short time spans, our sensitivity analysis is motivated by the assumption that method misspecification, not statistical noise, drives error in treatment effect estimates. As a result, we caution against interpretation of our analysis as a statistical inference procedure, although for a particular choice of misspecification error metric we can view our procedure as a geometric motivation for residual-based randomization tests. We demonstrate the value of our sensitivity analysis in the context of three canonical comparative case studies for which the SC method has been used. 

One potential avenue for future study is incorporating statistical uncertainty into our sensitivity analysis framework. Given the difficulty of characterizing the treatment assignment mechanism in cases with a single treated unit, it would make the most sense to assume outcomes are stochastic and independent across units with bounded variance heterogeneity as in \cite{hagemann2020inference}. In this setting, we could measure misspecification error in terms of the bias of the ``pseudo-true'' SC weights computed by minimizing the expectation of the usual SC objective \citep{chernozhukov2018practical,cattaneo2019prediction}. If for simplicity we conditioned on pre-treatment outcomes as in \cite{cattaneo2019prediction} and assumed the magnitude of misspecification error was at most the placebo misspecification error of some percentage of the control units,\footnote{Such a perspective is reminiscent of the partial identification approach taken in \cite{rambachan2019honest} to allow for limited violations of the parallel trends assumption in the context of event studies} we could potentially develop a conditional prediction interval for the treated unit's outcome in a given period \citep{cattaneo2019prediction}. Of course, there is much more to be done to understand the viability (or lack thereof) of this general approach given the conceptual difficulty in measuring variability due to sampling in comparative case study settings, so we leave doing so to future work.

In conclusion, we hope that researchers will perform the sensitivity analysis outlined in Procedures \ref{alg:weight_space_sensitivity_analysis} and \ref{alg:general_sensitivity_analysis} as part of their future comparative case studies employing the SC method and visualize their results as in Figures \ref{fig:case_study_plots} and \ref{fig:case_study_compare}.

\pagebreak

\bibliography{references}

\pagebreak

\appendix

\section{Placebo Analysis of Procedures \ref{alg:weight_space_sensitivity_analysis} and \ref{alg:general_sensitivity_analysis}} \label{sec:placebo_analysis}

We perform our sensitivity analysis on each of the control units in the three case studies as an extension of the placebo analysis used to assess the performance of our procedure at the end of Section \ref{sec:case_studies}. We treat each control unit as the placebo treated unit and run Procedure \ref{alg:general_sensitivity_analysis} under the three misspecification error metrics discussed in Section \ref{subsec:case_studies_revisited}. For each placebo treated unit, our procedure returns the minimum percentile rank of the placebo control units' misspecification errors at which a zero treatment effect is in the range of effects plausible under the allowed misspecification error. Within each case study, we can vary the level of acceptable misspecification error by choosing a different percentile rank cutoff. At each proposed cutoff, we generate bounds on the treatment effect and observe the share of control units for which we correctly include the zero treatment effect.

\begin{figure}[p]
\centering
\captionsetup{width=\textwidth}
\begin{subfigure}{\textwidth}
\centering
\includegraphics[width=0.55\textwidth]{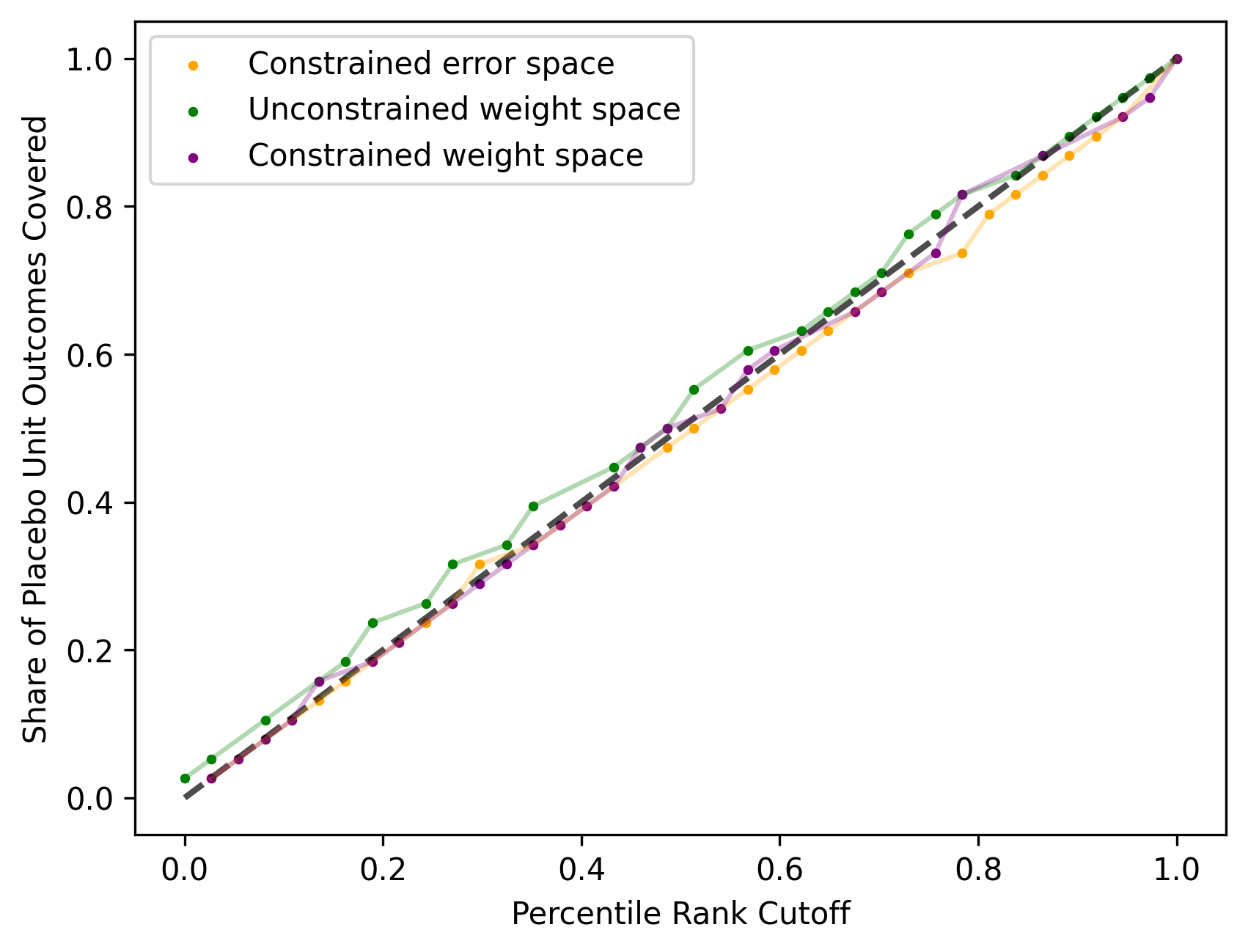}
\caption{Effect of California's Tobacco Control Program}
\label{fig:california_placebo_plot}
\end{subfigure}
\begin{subfigure}{\textwidth}
\centering
\includegraphics[width=0.55\textwidth]{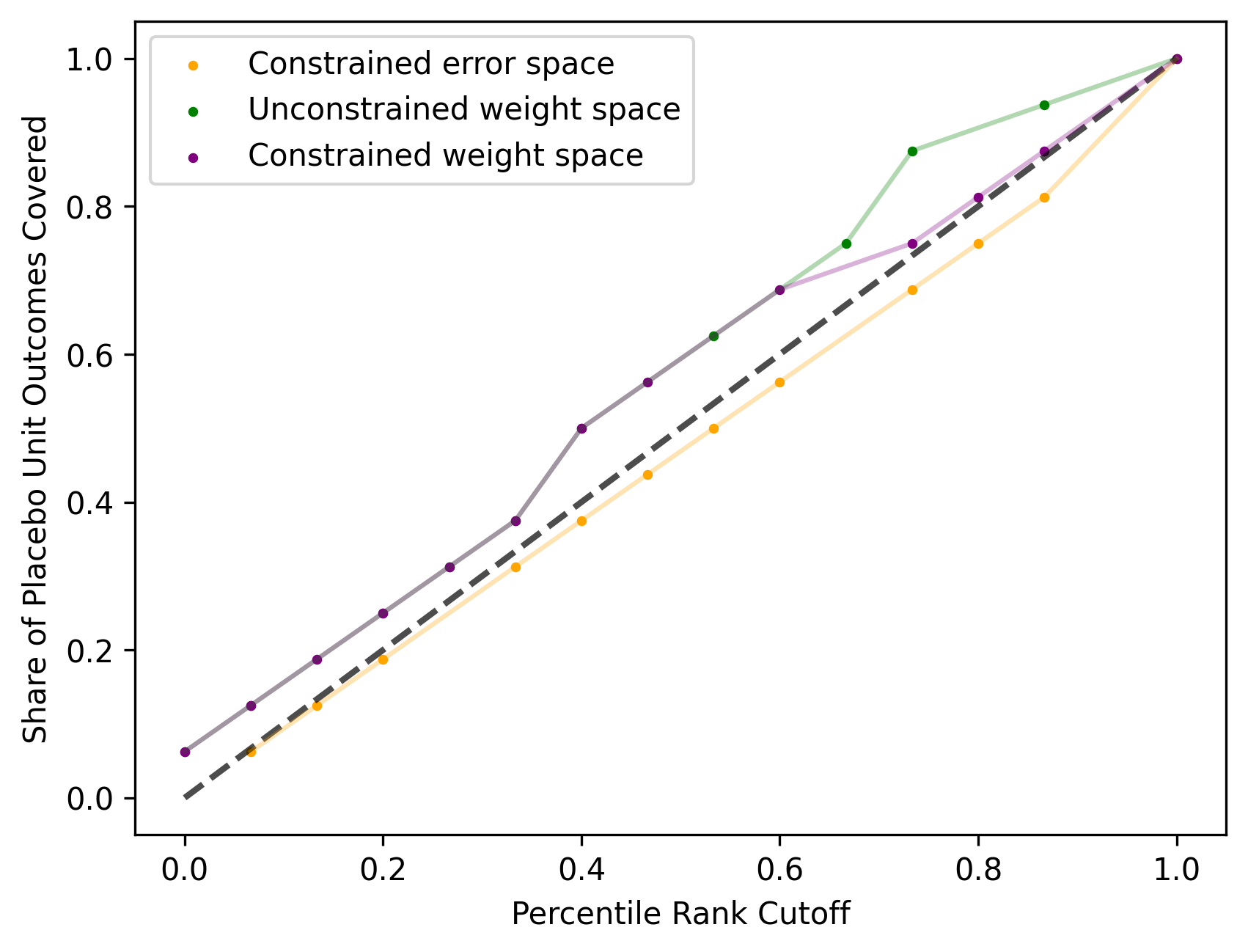}
\caption{Effect of German Reunification on GDP}
\label{fig:german_placebo_plot}
\end{subfigure}
\begin{subfigure}{\textwidth}
\centering
\includegraphics[width=0.55\textwidth]{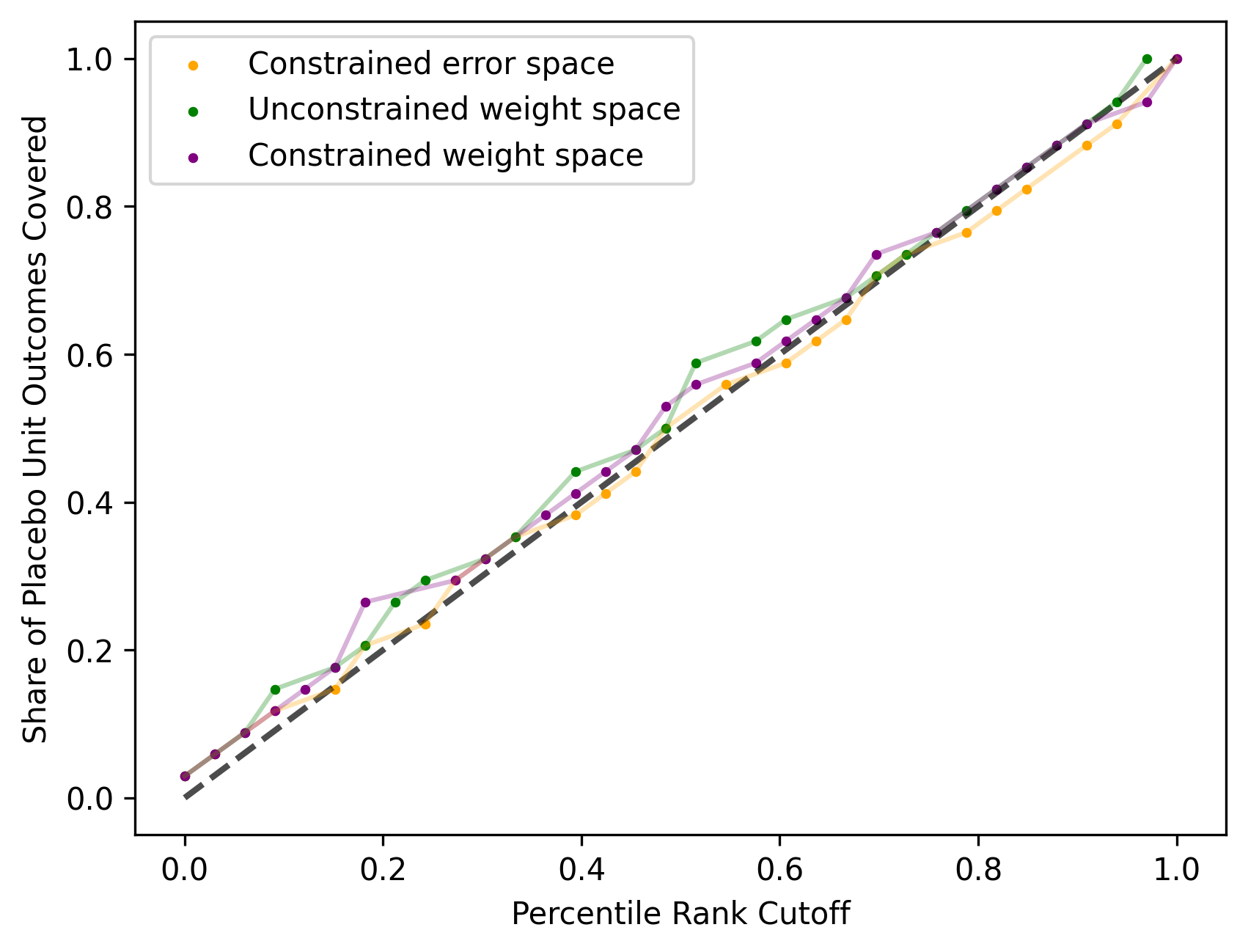}
\caption{Effect of Mariel Boatlift on Low-Income Wages}
\label{fig:mariel_boatlift_placebo_plot}
\end{subfigure}
\caption{We plot the results from our placebo analysis of Procedure \ref{alg:general_sensitivity_analysis} using three different misspecification error metrics in each of the three case studies, as in Section \ref{subsec:case_studies_revisited}. On the $x$-axis, we vary the percentile rank cutoff that determines the width of the bounds generated by our procedure and on the $y$-axis we show the share of placebo treated units for which those bounds correctly include a zero treatment effect. The dashed black 45-degree line demarcates $p$ percent of placebo units' true outcomes being covered with a misspecification error cutoff at the $p$th percentile rank.}
\label{fig:case_study_placebo_plots}
\end{figure}

In Figure \ref{fig:case_study_placebo_plots}, we report the results of this placebo exercise by plotting the share of control units for which our procedure generates bounds that correctly contain zero treatment effect under each possible percentile rank cutoff. Across case studies and misspecification error metrics, a $p$th percentile rank cutoff is associated with correct predictions for around $p$ percent of the control units. This direct correspondence between percentile rank cutoff and the coverage of placebo units' outcomes is to be expected given the design of our placebo analysis. 

When our sensitivity analysis is performed on the true treated unit, the set of placebo misspecification errors is calculated when, for each control unit $j$, we calculate the distance (in a chosen metric) between the canonical SC estimate using the remaining $J - 1$ control units and the closest weights that correctly predict the target outcome. If we were to plot the share of control units covered at each percentile rank of the misspecification error distribution generated in our analysis of the true treated unit, we would exactly recover the 45-degree line. However, when we perform the analysis on control unit $j$ as a placebo treated unit, the set of misspecification errors is constructed by creating placebo SC weights for the remaining placebo control units without unit $j$, that is, using $J - 2$ control units. Redefining the set of misspecification errors without using placebo control unit $j$ can yield slightly different SC estimates that generate the deviations seen in each of panel of Figure \ref{fig:case_study_placebo_plots}. 

In Figure \ref{fig:case_study_placebo_plots}, we also observe that the unconstrained weight space metric yields placebo coverage rates that always lie weakly above the 45-degree line. While this result is not theoretically guaranteed, we can see why we may expect such a phenomenon. Consider the placebo analysis of control unit $j$. Using the remaining $J-1$ control units, the closed form solution for the observable misspecification error under the unconstrained weight space metric is given in Equation \eqref{eq:min_dist_expression}: the ratio of the absolute SC residual to the $\ell_2$-norm of the vector of $J-1$ control unit outcomes, $B_0^{(j)} = {\abs{\hat{R}_{jT^*}^\text{sc}}}/{\norm{\mathbf{Y}_{(-j)T^*}}_2}$. Placebo treated unit $j$'s misspecification error will be compared to the distribution of misspecification errors of the remaining $J-1$ placebo control units, calculated as each placebo control unit $k$'s ratio of the absolute residual from predicting its post-treatment SC estimate using the remaining $J-2$ control units to the $\ell_2$-norm of the vector of $J-2$ remaining control unit outcomes, $d_2(\mathbf{w}_\text{sc}^{(k)}, \mathcal{W}_k^*) = {\abs{\hat{R}_{kT^*}^\text{sc}}}/{\norm{\mathbf{Y}_{(-j, -k)T^*}}_2}$. Because most SC estimates are sparse and thus do not put any weight on the placebo treated unit $j$ anyway, the residuals will not change much whether $J-1$ or $J-2$ placebo control units are used to construct SC estimates, in which case  the primary difference in misspecification error will be driven by the reduction in the norm of the vector of control unit outcomes with $J-2$ units instead of $J-1$ units, $\norm{\mathbf{Y}_{(-j, -k)T^*}}_2 \leq \norm{\mathbf{Y}_{(-j)T^*}}_2$. Therefore, it is reasonable to expect that the treatment effect bounds for control unit $j$ will include a zero treatment effect at a weakly lower percentile rank in the placebo analysis than in the sensitivity analysis of the true treated unit, which is why the unconstrained weight space placebo coverage rates lie weakly above the 45-degree line.

While there is not an observable pattern in the constrained weight space case, we do observe that the constrained error space metric yields placebo coverage rates that are mostly below the 45 degree line. Recall from Equation \ref{eq:def_m_j_err} and Procedure \ref{alg:general_sensitivity_analysis} that the constrained error space misspecification error \eqref{eq:min_dist_to_opt_weights_control_unit_general} is the ratio of the pre-treatment error from the best-fitting performing synthetic control pre-treatment that achieves perfect post-treatment accuracy to the minimum SC pre-treatment error. When placebo treated unit $j$'s misspecification error is compared to the errors of the remaining $J-1$ units that are fit with only $J-2$ placebo control units, both the numerator and denominator of the control unit misspecification errors weakly increase as pre-treatment fit can only get worse with fewer units. This behavior makes the relative magnitudes of control unit misspecification errors computed using $J-2$ control units to the equivalent errors computed using $J-1$ control units theoretically ambiguous. However, since the constrained error space metric placebo coverage rates tend to be below the 45-degree line, it must be that the misspecification errors with $J-1$ units are typically smaller than with $J-2$ units. This suggests that the addition of a control unit in the donor pool tends to decrease pre-treatment fit error constrained to perform well in the post-treatment period of interest less than it decreases the minimum achievable pre-treatment error. 

\section{Handling Non-Uniqueness of the SC Estimator}\label{sec:mult_solns_appendix}

As discussed in Footnote \ref{foot:mult_solns} in Section \ref{subsec:notation}, it is possible that $\mathbf{x}_1$ could lie in the convex hull of the columns of $X_0$, in which case the optimization problem \eqref{eq:basic_synth_contr_method} that defines the SC estimator could have multiple or even infinite solutions. In this case, the sensitivity analysis described in Section \ref{sec:sensitivty_analysis} would understate the impact of misspecification error on treatment effect estimates because it does not account for the multiplicity of valid SC estimators that could result from solving \eqref{eq:basic_synth_contr_method}. Thankfully, we can apply the generalized sensitivity analysis described in Section \ref{subsec:gen_sensitivity_analysis} with an appropriate choice of misspecification error metric $m_j$ to allow for a weight-space-based sensitivity analysis that can account for non-uniqueness.

In particular, we can choose $m_j^\text{wt,mult}(\mathbf{w})$ to measure the distance of $\mathbf{w}$ to the \emph{closest} (in $\ell_2$ distance) SC weights that solve \eqref{eq:basic_synth_contr_method}. Formally, we let $$\mathcal{W}_\text{sc}^{(j)} \coloneqq \argmin_{\mathbf{w} \in \R^{J - \ind{j \neq 1}}} \left\{\normnofit{\mathbf{x}_j - X_{(-j)}\mathbf{w}}_2 + \psi_{\Delta_{J - \ind{j \neq 1}}}(\mathbf{w})\right\}$$
denote the set of optimal solutions to \eqref{eq:basic_synth_contr_method} with penalty term $\psi$ defined in \eqref{eq:constraint_penality} and define
\begin{equation*}
	m_j^\text{wt,mult}(\mathbf{w}) \coloneqq \inf_{\tilde{\mathbf{w}} \in \R^{J - \ind{j \neq 1}}}\left\{\norm{\tilde{\mathbf{w}} - \mathbf{w}}_2 + \psi_{\mathcal{W}_\text{sc}^{(j)}}(\tilde{\mathbf{w}}) \right\}.
\end{equation*}
If we apply our generalized sensitivity analysis using misspecification error metric $m_j^\text{wt,mult}$, the resulting treatment effect bounds will capture the impact of both misspecification error and potential non-uniqueness of the SC estimator because, in effect, the solutions to \eqref{eq:min_dist_to_opt_weights_control_unit_general} and \eqref{eq:bounds_on_cntrfct_control_outcomes_general} will range over all weights that lie within $d_{m_j}(\mathbf{w}_\text{sc}^{(j)}, \mathcal{W}_j^*)$ of \emph{some} valid SC estimator, not just the particular SC estimator returned by the estimation procedure.

Note that since $\mathcal{W}_\text{sc}^{(j)}$ is defined as the solution set of a convex optimization problem and must therefore be a convex set (see Section 4.2.1 of \cite{boyd2004convex}), and since the infimum of a convex function in one of its arguments over a convex set must be convex in its remaining arguments (again, see \cite{boyd2004convex}), $m_j^\text{wt,mult}$ must convex. However, \eqref{eq:min_dist_to_opt_weights_control_unit_general} and \eqref{eq:bounds_on_cntrfct_control_outcomes_general} with $m_j^\text{wt,mult}$ directly plugged in are not formulated in manners that are particularly amenable to computation. Instead, we can write \eqref{eq:min_dist_to_opt_weights_control_unit_general} using misspecification error metric $m_j^\text{wt,mult}$ in a form that is more directly solvable as follows:
\begin{equation}\label{eq:min_dist_to_opt_weights_control_unit_non_unique}
\begin{aligned}
	d_{m_j}(\mathbf{w}_\text{sc}^{(j)}, \mathcal{W}^*_j) \coloneqq \inf_{\tilde{\mathbf{w}}, \mathbf{w} \in \R^{J-1}}&~ \norm{\tilde{\mathbf{w}} - \mathbf{w}}_2 \\
		\suchthat&~ \mathbf{Y}_{(-j)T^*}^T\mathbf{w} = Y_{jT^*} ~~\left(\Leftrightarrow \mathbf{w} \in \mathcal{W}_j^*\right) \\
		&~ \normnofit{\mathbf{x}_j - X_{(-j)}\tilde{\mathbf{w}}}_2 \leq \normnofit{\mathbf{x}_j - X_{(-j)}\mathbf{w}_\text{sc}^{(j)}}_2 \\
		&~ \ones^T\tilde{\mathbf{w}} = 1 \\
		&~ \tilde{\mathbf{w}} \geq \zeros.
\end{aligned}
\end{equation}
Similarly, we can write the optimization problems in \eqref{eq:bounds_on_cntrfct_control_outcomes_general} to facilitate use of common convex solvers as follows:
\begin{equation}\label{eq:bounds_on_cntrfct_control_outcomes_non_unique}
\begin{aligned}
	Y_{1T^*}^{B_j,-}(0) \coloneqq \inf_{\tilde{\mathbf{w}}, \mathbf{w} \in \R^{J-1}}&~ \mathbf{Y}_{0T^*}^T\mathbf{w} \\
		\suchthat&~ \norm{\tilde{\mathbf{w}} - \mathbf{w}}_2 \leq d_{m_j}(\mathbf{w}_\text{sc}^{(j)}, \mathcal{W}^*_j) \\
		&~ \normnofit{\mathbf{x}_j - X_{(-j)}\tilde{\mathbf{w}}}_2 \leq \normnofit{\mathbf{x}_j - X_{(-j)}\mathbf{w}_\text{sc}^{(j)}}_2 \\
		&~ \ones^T\tilde{\mathbf{w}} = 1 \\
		&~ \tilde{\mathbf{w}} \geq \zeros,
\end{aligned}
\end{equation}
and $Y_{1T^*}^{B_j,+}(0)$ is defined similarly. Finally, we note that generalizing the constrained weight space sensitivity analysis discussed in Section \ref{subsec:case_studies_revisited} to allow for non-uniqueness is essentially the same as the process described above with additional constraints $\ones^T\mathbf{w} = 1$ and $\mathbf{w} \geq \zeros$ added to \eqref{eq:min_dist_to_opt_weights_control_unit_non_unique} and \eqref{eq:bounds_on_cntrfct_control_outcomes_non_unique}.

\section{Other Generalizations}\label{sec:other_generalizations}

\subsection{Other Contrasts}\label{subsec:other_contrasts}

While the treatment effect $\tau_{T^*}$ in period $T^*$ is a natural estimand in comparative case study settings, researchers are often interested in other linear contrasts of outcomes like the average treatment effect across all post-treatment periods or the effect on the average slope of the treated unit's outcome path. Our sensitivity analysis can be extended naturally to assess the robustness of synthetic control-based estimates of these alternative estimands.

Let $\mathbf{Y}_{j,T_0:T}(d) \coloneqq (Y_{jT_0}(0), \dotsc, Y_{jT}(0))^T$ denote the vector containing unit $j$'s potential outcomes under treatment arm $d$ in each of the $T - T_0$ post-treatment periods, and suppose we are interested in assessing the robustness of an estimand $\tau_c$ parameterized by the vector $c \coloneqq (c(0), c(1))^T \in \R^{2(T - T_0)}$:
\begin{equation*}
	\tau_c \coloneqq c^T\bmat{Y_{1, T_0:T}(1) \\ Y_{1, T_0:T}(0)} = c(1)^TY_{1, T_0:T}(1) + c(0)^TY_{1, T_0:T}(0).
\end{equation*}
For example, if we use the contrast vector $c_{T_*}$ defined entrywise as $$[c_{T_*}(d)]_t \coloneqq \ind{t = T^*}(d - (1-d)),$$ we recover the treatment effect in period $T^*$ studied in Section \ref{sec:sensitivty_analysis}, $\tau_{T^*} = \tau_{c_{T_*}}$. If we instead use $c_\text{avg}$ defined entrywise as $$[c_\text{avg}(d)]_t \coloneqq \frac{1}{T - T_0}(d - (1-d)),$$ we recover the average treatment effect $\tau_{c_\text{avg}}$ across the $T - T_0$ post-treatment periods. Using $c_\text{slo}$ defined entrywise as $$[c_\text{slo}(d)]_t \coloneqq \frac{1}{T - T_0}(\ind{t = T} - \ind{T = T_0})(d - (1-d))$$ yields the effect $\tau_{c_\text{slo}}$ on the average slope of the treated unit's outcome path, since the sums in the average slopes telescope:
\begin{align*}
	\tau_{c_\text{slo}} &\coloneqq \frac{1}{T - T_0}\sum_{t = T_0}^{T-1} (Y_{1(t+1)}(1) - Y_{1t}(1)) - \frac{1}{T - T_0}\sum_{t = T_0}^{T-1} (Y_{1(t+1)}(0) - Y_{1t}(0)) \\
	&= \frac{1}{T - T_0}\left\{\left[Y_{1T}(1) - Y_{1T_0}(1)\right] - \left[Y_{1T}(0) - Y_{1T_0}(0)\right]\right\}
\end{align*}

Next, let $\mathbf{Y}_{j,T_0:T} \coloneqq (Y_{jT_0}(D_{jT_0}), \dotsc, Y_{jT}(D_{jT}))^T$ be the vector of unit $j$'s observed post-treatment outcomes, let $\mathbf{Y}_0$ be the $(T - T_0) \times J$ matrix of control units' post-treatment outcomes, where the $j$th column of $\mathbf{Y}_0$ is $\mathbf{Y}_{j+1, T_0:T}$, and let $\mathbf{Y}_{-j}$ denote the matrix $\mathbf{Y}_0$ with its $j$th column deleted. Then once we have chosen a contrast $c$, we can write the synthetic control estimate of $\tau_c$ for the treated unit as follows:
\begin{equation*}
	\hat{\tau}_c^\text{sc} \coloneqq c(1)^T\mathbf{Y}_{1, T_0:T} + c(0)^T\hat{\mathbf{Y}}_{1, T_0:T} = c(1)^T\mathbf{Y}_{1, T_0:T} + c(0)^T\mathbf{Y}_0\mathbf{w}_\text{sc}.
\end{equation*}
Similarly, we can write the placebo treatment effect for the $j$th control unit using the other $J-1$ control units as the donor pool as follows:
\begin{equation*}
	\hat{\tau}_c^{(j),\text{sc}} \coloneqq c(1)^T\mathbf{Y}_{j, T_0:T} + c(0)^T\hat{\mathbf{Y}}_{j, T_0:T} = c(1)^T\mathbf{Y}_{j, T_0:T} + c(0)^T\mathbf{Y}_{-j}\mathbf{w}_\text{sc}^{(j)}
\end{equation*}

Given this characterization of $\hat{\tau}_c^\text{sc}$, modifying the general procedure described in in Section \ref{subsec:misspec_generalize} is relatively straightforward. First, we replace the constraints $\mathbf{Y}_{(-j)T^*}^T\mathbf{w} = Y_{jT^*}$ and $\mathbf{Y}_{0T^*}^T\mathbf{w} = Y_{1T^*}$ requiring perfect post-treatment accuracy in period $T^*$ in the optimization problems \eqref{eq:min_dist_to_opt_weights_control_unit_general} and \eqref{eq:nullifying_distance_comp_def_general} with the constraints $c(0)^T\mathbf{Y}_{j, T_0:T} = c(0)^T\mathbf{Y}_{-j}\mathbf{w}$ and $c(0)^T\mathbf{Y}_{1, T_0:T} = c(0)^T\mathbf{Y}_{0}\mathbf{w}$, which is equivalent to requiring correct treatment effect estimation for control unit $j$ in the case of \eqref{eq:min_dist_to_opt_weights_control_unit_general} and for the treated unit under the assumption of no effect in \eqref{eq:nullifying_distance_comp_def_general}. The definition of $d_{m^{(j)}}(\mathbf{w}_\text{sc}^{(j)}, \mathcal{W}^*_j)$ should also be updated accordingly. Next, we replace the computations of the bounds on $Y_{1T^*}(0)$ in \eqref{eq:bounds_on_cntrfct_control_outcomes_general} with the following bounds on the component of the treatment effect that depends on counterfactual control outcomes $Y_{1,T_0:T}(0)$:
\begin{equation*}\label{eq:bounds_on_cntrfct_control_outcomes_alt_contrast}
	\begin{aligned}
		\mu_{1}^{B_j,-}(0) &\coloneqq \min_{\mathbf{w} \in \R^J} \left\{c(0)^T\mathbf{Y}_{-j}\mathbf{w} 
		~\setst~ m^{(j)}(\mathbf{w}) \leq d_{m^{(j)}}(\mathbf{w}_\text{sc}^{(j)}, \mathcal{W}^*_j)\right\} \\
		\mu_{1}^{B_j,+}(0) &\coloneqq \max_{\mathbf{w} \in \R^J} \left\{c(0)^T\mathbf{Y}_{-j}\mathbf{w}
		~\setst~ m^{(j)}(\mathbf{w}) \leq d_{m^{(j)}}(\mathbf{w}_\text{sc}^{(j)}, \mathcal{W}^*_j)\right\}
	\end{aligned}
	\end{equation*}
Finally, we replace the bounds in \eqref{eq:def_T_B_j_bounds_gen} with the following bounds on $\tau_c$:
\begin{equation*}\label{eq:def_T_B_j_bounds_alt_contrast}
	\begin{aligned}
		\mathcal{T}^{B_j}_c &\coloneqq \left[c(1)^T\mathbf{Y}_{1, T_0:T} + \mu_{1}^{B_j,-}(0), c(1)^T\mathbf{Y}_{1, T_0:T} + \mu_{1}^{B_j,+}(0)\right]
	\end{aligned}
	\end{equation*}

\subsection{Other Panel Data Methods}\label{subsec:lpes}

The outcomes-based SC method described in Section \ref{subsec:notation} is by no means the only method for generating counterfactual predictions in comparative case study settings. Besides the classic Difference-in-Differences estimator \citep{bertrand2004} and SC estimators that incorporate other pre-treatment covariates \citep{abadie2015comparative}, a whole suite of methods for panel data inspired by the SC method have been proposed in the past decade, including but not limited to the estimators proposed in \cite{doudchenko2016balancing}, \cite{chernozhukov2017exact}, \cite{ben2018augmented}, \cite{crest2018penalized}, \cite{arkhangelsky2020synthetic}, and \cite{kellogg2020combining}.

As has been noted in \cite{doudchenko2016balancing}, \cite{chernozhukov2017exact}, and \cite{cattaneo2019prediction} among others, we can write many of these alternative treatment effect estimators for panel data as affine functions of the control units' post-treatment outcomes $\mathbf{Y}_{0T^*}$ in period $T^*$:
\begin{equation}\label{eq:gen_treat_eff_estimate}
\hat{\tau}_{T^*} \coloneqq Y_{1T^*} - \left(\hat{\mu} + \mathbf{Y}_{0T^*}^T\hat{\mathbf{w}}\right) = Y_{1T^*} - \bmat{1 & \mathbf{Y}_{0T^*}}\bmat{\hat{\mu} \\ \hat{\mathbf{w}}},
\end{equation}
where we now allow for an intercept term $\hat{\mu}$ in addition to weights on the control units' post-treatment outcomes. As indicated by the second equality in \eqref{eq:gen_treat_eff_estimate}, to allow for an intercept term, we can simply add an extra ``control unit'' to the donor pool with ones as all of its outcomes, so we assume we include such an intercept unit and omit the explicit intercept term in what follows.

Once we take this more general perspective, we can see that Procedure \ref{alg:weight_space_sensitivity_analysis} easily generalizes to accommodate alternative policy evaluation methods that generate treatment effect estimates as in \eqref{eq:gen_treat_eff_estimate}, since all the procedure requires are the weights on units' post-treatment outcomes that generate counterfactual outcome predictions. Specifically, the only difference is that instead of using the SC method to generate the weights used to compute the residuals in steps \ref{alg:weight_space_placebo_synth_step} and \ref{alg:weight_space_nullifying_error_step}, we use the weights outputted by some alternative policy evaluation method.

Further, many of the methods listed above can be described as choosing weights to solve a particular instance of the following general convex program:
\begin{equation}\label{eq:gen_method_opt_prob}
\begin{aligned}
	J_{\hat{V},r,C}(\mathbf{w}) &\coloneqq (\mathbf{x}_1 - X_0\mathbf{w})^T\hat{V}(\mathbf{x}_1 - X_0\mathbf{w}) + r(\mathbf{w}) + \psi_C(\mathbf{w}), \\
	\hat{\mathbf{w}} &\coloneqq \argmin_{\mathbf{w} \in \R^{J+1}} J_{\hat{V},r,C}(\mathbf{w})
\end{aligned}
\end{equation}
where $C$ is a convex set, $\psi_C$ is a penalty term as defined in \eqref{eq:constraint_penality} to ensure $\hat{\mathbf{w}} \in C$, $\hat{V}$ is a weighting matrix that can be chosen in a potentially data-driven manner, $r \colon \R^J \rightarrow [0, \infty)$ is a convex penalty term that regularizes the weights $\mathbf{w}$ in some fashion, the columns of $X_0$ can contain additional pre-treatment covariates beyond control units' pre-treatment outcomes, and we include an additional first column in $X_0$ containing ones in all the rows corresponding to pre-treatment outcomes and zeros in the other rows. For brevity, we leave the reader to see how the methods listed above can be written in the form of \eqref{eq:gen_method_opt_prob} in the papers introducing them.

Given this common characterization of many alternative policy evaluation methods, we can see that for an appropriately defined misspecification error metric $m_j(\mathbf{w})$ that measures the misspecification error incurred by an alternative policy evaluation method's weights relative to the weights $\mathbf{w}$, the general Procedure \ref{alg:general_sensitivity_analysis} can be used without modification to assess the robustness of treatment effect estimates outputted by that alternative policy evaluation method. One particular misspecification error metric of interest is an analogue of the constrained error space metric $m^\text{err}_j$ defined in Section \ref{subsec:gen_sensitivity_analysis} corresponding to an alternative policy evaluation method defined by particular choices of $\hat{V}$, $r$, and $C$:
\begin{equation*}
m^\text{gen,err}_j(\mathbf{w}) = \frac{J_{\hat{V},r,C}(\mathbf{w})}{\min_{\tilde{\mathbf{w}} \in \R^{J - \ind{j \neq 1}}} J_{\hat{V},r,C}(\tilde{\mathbf{w}})} - 1.
\end{equation*}

\end{document}